\documentclass[fleqn,usenatbib]{mnras}

\usepackage{newtxtext,newtxmath}

\usepackage[T1]{fontenc}

\DeclareRobustCommand{\VAN}[3]{#2}
\let\VANthebibliography\thebibliography
\def\thebibliography{\DeclareRobustCommand{\VAN}[3]{##3}\VANthebibliography}

\usepackage{graphicx}	
\usepackage{amsmath}	
\usepackage{newtxtext,newtxmath}
\usepackage{multirow}

\title[Optical emission from cooling supernova shocks]{How do supernova remnants cool? - I. Morphology, optical emission lines, and shocks}

\author[E. I. Makarenko et al.]{
Ekaterina I. Makarenko$^{1}$\thanks{E-mail: makarenko@ph1.uni-koeln.de},
Stefanie Walch$^{1}$,
Seamus D. Clarke$^{2}$, Daniel Seifried$^{1}$, \\ \\ {\LARGE \rm Thorsten Naab$^{3}$, Pierre C. N{\"u}rnberger$^{1}$, Tim-Eric Rathjen$^{1}$} 
\\
$^{1}$I. Physikalisches Institut,  Universit{\"a}t zu K{\"o}ln, Z{\"u}lpicher Str. 77, D-50937 K{\"o}ln, Germany\\
$^{2}$Institute of Astronomy and Astrophysics, Academia Sinica, No. 1, Sec. 4, Roosevelt Rd., Taipei 10617, Taiwan\\
$^{3}$Max Planck Institute for Astrophysics, Karl-Schwarzschild-Str. 1, 85748 Garching, Germany\\
}

\date{Accepted 2023 May 8. Received 2023 April 3; in original form 2023 April 3}

\pubyear{2023}

\begin{document}
\label{firstpage}
\pagerange{\pageref{firstpage}--\pageref{lastpage}}
\maketitle

\begin{abstract}
Supernovae (SNe) inject $\sim 10^{51}$ erg in the interstellar medium, thereby shocking and heating the gas. 
A substantial fraction of this energy is later lost via radiative cooling. We present a post-processing module for the {\sc FLASH} code to calculate the cooling radiation from shock-heated gas using collisional excitation data from {\sc MAPPINGS V}. 
When applying this tool to a simulated SN remnant (SNR), we find that most energy is emitted in the EUV. 
However, optical emission lines ($[$O III$]$, $[$N II$]$, $[$S II$]$, H${\alpha}$, H${\beta}$) are usually best observable. 
Our shock detection scheme shows that [S II] and [N II] emissions arise from the thin shell surrounding the SNR, while [O III], H$\rm \alpha$, and H$\rm \beta$ originate from the volume-filling hot gas inside the SNR bubble. 
We find that the optical emission lines are affected by the SNR's complex structure and its projection onto the plane of the sky because the escaping line luminosity can be reduced by 10 -- 80\% due to absorption along the line-of-sight.
Additionally, the subtraction of contaminating background radiation is required for the correct classification of an SNR on the oxygen or sulphur BPT diagrams.
The electron temperature and density obtained from our synthetic observations match well with the simulation but are very sensitive to the assumed metallicity. 
\end{abstract}




\section{Introduction}\label{sec:intro}

Massive stars ($> 8 \mathrm{M_{\odot}}$) have a short lifetime (only a few million years) and can end their lives as supernovae (SNe) surrounded by the molecular environment in which they formed.
As such, about 10 to 20 per cent of all SN remnants (SNRs) are estimated to interact with dense molecular clouds (MCs) \citep{Hewit&Yusef-Zadeh2009}.
Although the interaction of SNRs and the turbulent interstellar medium (ISM) is difficult to treat both analytically and numerically \citep{Haid16}, a solid analytical description of the evolution of SNRs in homogeneous media has long been established \citep{Sedov59, Chevalier77, Ostriker88}.
The density of the gas hit by the expanding SN shock dramatically affects the properties of the shock itself, such as the velocity and temperature of the shock \citep[e.g.][]{McKee77, Slane15, Chandra18}. 
The SNR shock also affects the molecular gas, heating and ionising it, and driving turbulence.
Recently, several authors have studied the evolution of SNRs interacting with the turbulent, multi-phase ISM in 3D simulations \citep[e.g.][]{deAvillez&Breitschwerdt2005, Gatto2015, Iffrig15, Walch15, Walch&Naab2015, Martizzi16, Seifried18, Zhang2019, Steinwandel2020}.
Current simulations of the multi-phase ISM show that only 10 to 20 per cent of all SNRs explode in a high-density environment due to SN clustering \citep{Hu2017, Gatto2017, Rathjen2021, Hislop2022}. This is in agreement with observational estimates by e.g. \citet{Hewit&Yusef-Zadeh2009}.

Observationally, SNRs can be studied across different energy bands due to their cooling radiation that covers the entire wavelength range \citep{Chandra18}.
Most SNRs are observed at radio wavelengths due to their non-thermal synchrotron emission \citep[e.g.][]{Blandford82}.
In an early evolutionary stage (Sedov-Taylor (ST) stage, typically $1-2\times 10^4$ years after the SN event), SNRs may be observed in the X-ray as the hot, shocked plasma cools \citep{Sasaki2004_ctb109_xmm, Sasaki2013_ctb109chandra, Slane15}.
In the later transition (TR) phase and in the pressure-driven snowplough (PDS) phase, when the dense, shocked gas cools, one may observe SNRs in the UV and Optical \citep{Dopita84}.
However, these observations are rarer because of extinction at optical wavelengths and because significant optical emission only occurs when high-density gas is shocked, i.e. when an SNR encounters a dense MC.
In Green's catalogue \citep{Green19} only about 20\% of the SNRs have optical counterparts. 

When observable, optical emission line diagnostics allow us to determine the evolutionary stage of the SNR and to analyse the physical parameters that define its environment.
Optical emission tends to originate from the cooling and recombination zone directly downstream of the shock front itself, which is why it can produce a variety of diagnostics for both the material encountered and the physical processes involved \citep{Dopita84, Blair85}.
The morphology of the line emission is typically filamentary or clumpy, possibly revealing the distribution of the dense gas (as predicted by, e.g. \citealt{McKee75} and observed by, e.g. \citealt{SC18, Mavromatakis01, Boumis09}).
The radiation signatures of shock waves can be highly complex, so synthetic observations should take into account the importance of the specific geometric projection \citep{Hester1987}.

Using well-observable optical emission line ratios, Active Galactic Nuclei (AGN), SNRs, and H\textsc{ii} regions can be classified by their main excitation mechanism.
For example, when observing $[$N II] ($\lambda 6583$), $[$O III] ($\lambda 5007$), $\mathrm{H{\alpha}}$, and $\mathrm{H{\beta}}$ lines, one can build emission line-ratio diagnostic tool, the so-called BPT diagram \citep{Baldwin81, Kauffmann2003, Kewley2019_diagn}.
On the BPT diagram, one can distinguish between the shock-dominated line ratios typically found in AGN-host galaxies (typically located in the upper right part of the BPT diagram, where both $[$O III] ($\lambda 5007$)/$\mathrm{H{\beta}}$ and $[$N II] ($\lambda 6583$)/$\mathrm{H{\alpha}}$ are elevated \citealt{Stasinska2006, Ho2008, Pagotto2020}), star-forming galaxies (typically found in the lower left part, where both $[$O III] ($\lambda 5007$)/$\mathrm{H{\beta}}$ as well as $[$N II] ($\lambda 6583$)/$\mathrm{H{\alpha}}$ are low; \citealt{Sanchez2015}), and the mixture of different ionisation mechanisms in between these two regimes where SNRs are typically found.
The exact position of the dividing line between the star formation (or H\textsc{ii} region)-dominated regime and the shock-dominated regime is still debated \citep{Kewley2001, Kauffmann2003, Herpich2006, Kewley2019} but the trends in the ionisation conditions across the BPT diagram are well known.

In fact, there have already been attempts to reproduce observations of SNRs from simulations \citep{Potter2014, Toledo-Roy14, Bolte2015, Orlando2019, Derlopa20}.
These are focused on either the remnant kinematics or the morphology.
Our analysis is inspired by the observations and analysis of SNRs in our Galaxy in different optical filters (lines) \citep[see, e.g. ][]{Fesen1985, Mavr2003, Mavr2004, Boumis2005}.
In this paper, we focus on optical line emission maps that trace the evolution of an SNR interacting with a dense MC.
We continue the work and post-process a high-resolution three-dimensional simulation from \citet{Seifried18} with the {\sc MAPPINGS V} code \citep{Sutherland17}.
{\sc MAPPINGS V} allows one to calculate the cooling radiation assuming collisional ionisation equilibrium (CIE).
In this way, we obtain the cooling energy emitted in various optical lines: $[$O III], $[$N II], $[$S II], $\mathrm{H}_{\alpha}$, and $\mathrm{H}_{\beta}$.
With this information, it is possible to classify and study the parameters of the object, for example using the "classic" version of the BPT diagram ($[$N II] ($\lambda 6583$)/$\mathrm{H}_{\alpha}$ versus $[$O III] ($\lambda 5007$)/$\mathrm{H}_{\beta}$) as well as the Veilleux-Osterbrock (VO) diagram ($[\mathrm{SII}] (\lambda 6531)/\mathrm{H}_{\alpha}$ vs. $[\mathrm{OIII}] (\lambda 5007)/\mathrm{H}_ {\beta}$; \citealt{Veilleux1987}) to build a complete set of optical line diagnostics.
Preliminary work was already done in \citet{Makarenko2020}.

The paper is structured as follows. Section~\ref{sec:numer_mod} describes the numerical model, the post-processing with {\sc MAPPINGS V}, and how we calculate the synthetic emission maps.
In Section~\ref{sec:post-processing} we discuss the line emission in the context of the remnant's evolution.
Next, in Section~\ref{sec:optical_emiss}, we present the results, focusing on the procedure of background subtraction, the optical emission of the SNR: attenuation and radiative transfer of optical emission lines, the BPT diagrams, and we determine the parameters of the SN shocks.
In Section~\ref{sec:conclusions}, we present our conclusions.


\section{Numerical model}
\label{sec:numer_mod}
We use one high-resolution, magneto-hydrodynamical (MHD) simulation of \citet{Seifried18} (MC1 MHD) carried out with the adaptive mesh refinement (AMR), finite-volume code {\sc FLASH} 4.3 \citep[see][]{Fryxell00, Dubey08}.
This simulation is part of the SILCC project \citep{Walch15, Girichidis16, Gatto2017, Peters2017, Girichidis2018, Rathjen2021, Rathjen2022} and the SILCC-Zoom project \citep{Seifried17}.
In Section~\ref{sec:sim} we briefly describe the included physics and assumptions, while we refer to the previous papers for more detailed information.

\subsection{Simulations}
\label{sec:sim}

The SILCC-Zoom simulation initially follows the formation of a MC from the multi-phase ISM under solar neighbourhood conditions.
The full size of the simulation box is 500 pc $\times$ 500 pc $\times $ $\pm$5 kpc.
With a vertical stratification in the $z$-direction and an initial gas surface density of $\mathrm{\Sigma_{gas} = 10\; M_{\odot} \ pc^{-2}}$, this volume presents a small section of the galactic disk.

We follow the estimate of \citet{Draine78} for the strength of the interstellar radiation field (ISRF), i.e. $G_{\rm 0}$ = 1.7 times the Habing flux.
The radiative transfer of the diffuse ISRF is self-consistently modelled using the {\sc TreeRay/OpticalDepth} module \citep[see][for a detailed description of the method]{Wunsch18}. 

A non-equilibrium chemistry network based on NL97 \citet{NelsonLanger1997} is included. It follows 7 species (H, H$^+$, H$_2$, C$^+$, CO, O and free electrons) and is used to compute the gas heating and cooling processes \citep[for further details see][]{Glover2007, Glover2010, Walch15}.
Because molecular hydrogen forms on dust, we have a simple dust model: the dust-to-gas mass ratio is fixed (a ratio of 1:100 is used), but the dust temperature is calculated separately from the gas temperature, assuming the dust to be in thermal equilibrium.
We consider several heating processes, such as heating by the photo-electric effect associated with the absorption of the ISRF photons \citep[for more details see][]{Walch15}.
However, in the simulation presented here, the predominant source of heating is the strong shock associated with the SNR and the corresponding thermalisation of the rarefied gas within the remnant bubble. UV and optical lines originate from the warm and hot ionised gas within the heated bubble and near the bubble rim.
For the cooling of the high-temperature gas ($T\gtrsim 2 \times 10^{4}$~K) we use tabulated cooling functions from \citet{GnatFerland12} for solar metallicity (warm ionised medium (WIM) model; see below) assuming CIE.
The cooling of the cooler gas takes into account the local elemental abundances traced by the chemical network.
Our simulation originally assumed that all cooling radiation simply escapes to infinity (a typical assumption for 3D (magneto-)hydrodynamic simulations), which is in principle only valid if the surrounding medium is optically thin at the relevant wavelengths.
We will show later on (see Section~\ref{sec:tau}), that optical depth effects actually do play a role for the considered SNR.
This will be further discussed in future work. 

We use metallicities resembling the Milky Way disc near the solar neighbourhood, i.e. we use the WIM model of \citet{Seembach2000}.
We summarise the most important elements for our analysis in Table~\ref{tab:metalls}. The WIM abundance corresponds to the solar abundance by \citet{Howk1999} modified by dust as observed in warm diffuse clouds.
The abundances of nitrogen and oxygen adapted for the WIM are from \citet{MeyerNitrogen1997, MeyerOxygen1998}.
The magnetic field value is set to  B = 3~$\mu$G.

\begin{table}
    \centering
    \begin{tabular}{c|c|c|c|c}
        \hline
        Element & Dopita2005 A(X) & WIM A(X) & Solar A(X) & 2xSolar A(X)\\
        \hline
        N & 7.58 & 7.88 & 8.05 & 8.35 \\
        O & 8.44 & 8.50 & 8.93 & 9.23 \\
        S & 6.99 & 7.07 & 7.21 & 7.51 \\
        \hline
    \end{tabular}
    \caption{Metallicities for the main elements (column 1) whose ions are typically observable in the SNRs.
    The abundances are given as logarithmic values on a scale where A(X) = log (X/H) + 12.00 and increase from left to right.
    The Dopita2005 (column 2), default "solar abundance" (column 4), "2 solar " (column 5) in {\sc MAPPINGS V} are taken from \citet{Asplund2009}. The solar abundance "WIM" \citep{Seembach2000} (column 3) is what we use in our simulations.}
    \label{tab:metalls}
\end{table}

The simulation starts off by driving the multi-phase ISM with SNe for more than 10 Myr before following the formation of a MC in detail.
The zoom-in procedure uses the AMR capabilities of the {\sc Flash} code: within a sub-region of $\sim (100\;{\rm pc})^3$ we allow the code to adaptively refine down to a minimum cell size of 0.12~pc, while the rest of the 500~pc-scale computational domain is kept unchanged. 
After the forming MC is fully refined ($t_0 =$ 11.9 Myr) and a bit more evolved, we explode a single SN at time $t_\mathrm{SN} = t_0 +$ 1.53 Myr and at a distance of $d=25$~pc from the centre of mass of the cloud along the $x-$direction. 
The SN explosion is modelled by injecting a total amount of $E_{\rm SN}=10^{51}$~erg in the form of thermal energy within a spherical injection region of 4.5 grid cells, which corresponds to 11.7 pc in our case. At this resolution and injection radius, the Sedov-Taylor phase is well resolved, and therefore a thermal energy injection scheme is suitable \citep[e.g., see][]{Gatto2015, Kim2015}. 
For more details on the simulation see "MC1-MHD +25pc x" in \citet{Seifried17, Seifried18}. We note that for the analysis and emission line diagnostics presented in the following, we use a slightly smaller volume (65 pc $\times$ 76 pc $\times$ 68 pc around the explosion centre). We further note that we study a whole set of different simulations in a follow-up paper (Smirnova et al., in prep.).

We follow how the SNR expands and also interacts with the nearby dense MC over time $t_\mathrm{evol} = t - t_\mathrm{SN}$. Specifically, the evolution of the remnant in this paper is followed until $t_\mathrm{evol,fin}=$ 0.3 Myr. This is motivated by the following time scale estimates. The SNR transition time (TR) marks the end of the adiabatic Sedov-Taylor (ST) phase. Following the calculation of \citet{Haid16}, we obtain an SNR transition time of $t_\mathrm{TR} \approx 0.042$ Myr for an assumed uniform environmental number density of $n= 1$~cm$^{-3}$ of the SN explosion site.
The time at which the remnant would then transit to the pressure-driven snowplough (PDS) phase is $t_\mathrm{PDS} \approx 0.08$ Myr.
With these estimates, we may compare the optical emission obtained from synthetic and real observations.
It can also be seen that the time span of 0.3~Myr past explosion should be sufficient to trace the main part of the SNR evolution. We will further discuss these time scale estimates in Section~\ref{sec:SNR_times}.

\subsection{MAPPINGS V post-processing and cooling function}
\begin{figure}
	\includegraphics[trim={0.6cm 0.7cm 0.5cm 0.5cm},clip, width=1.0\columnwidth]{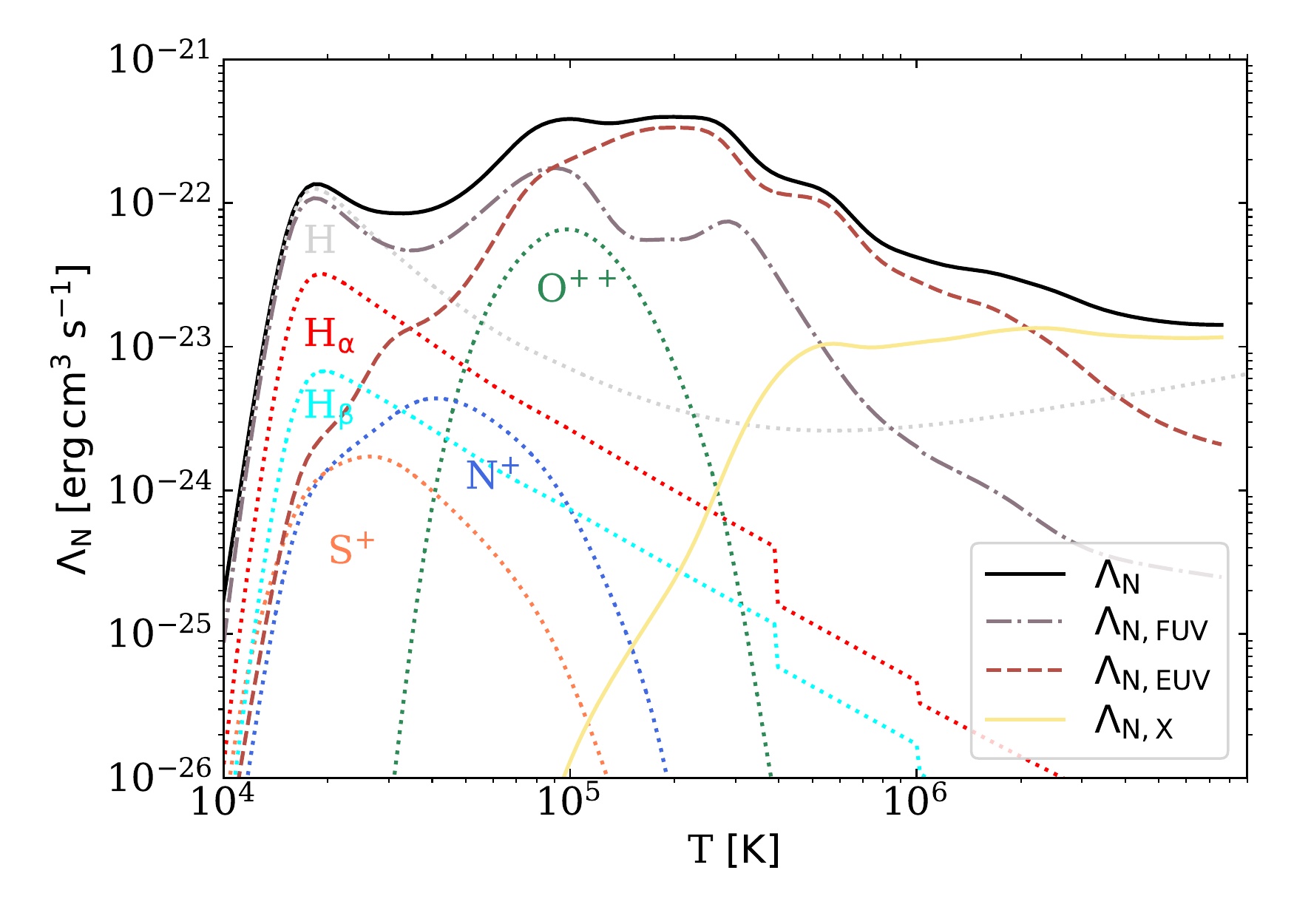}
    \caption{Normalised cooling function, i.e. emissivities $\Lambda_{\rm N}$ as a function of the gas temperature calculated using {\sc MAPPINGS V} for solar metallicity (WIM model; for details on the metal abundances see \citet{Seembach2000} and table~\ref{tab:metalls}).
    In addition to the total emissivity (black line), we also show the normalized cooling curves in three different energy bands: the far-ultraviolet (FUV) and optical ($E_{\rm FUV} <$ 13.6~eV), the ionizing EUV (13.6 eV $\le E_{\rm EUV} <$ 100~eV), and the X-ray ($E_{\rm X} \ge$ 100~eV).
    Further, we overplot the contributions of the ions used in the emission line diagnostics.}
    \label{fig:map_cc}
\end{figure}

While, during the course of the 3D simulation, the numerical scheme only works with a total cooling rate, we are interested to see what fraction of the energy lost by radiative cooling is emitted in the X-ray ($E_{\rm X} \ge$ 100~eV), ionising UV (EUV; 13.6 eV $\le E_{\rm EUV} <$ 100~eV), or the non-ionising UV (FUV) and Optical ($E_{\rm FUV} <$ 13.6~eV).
In particular, we also calculate the emission in the most bright optical emission lines such as [OIII], [NII], and [SII], which can be used for object diagnostics.

For this purpose, we use the {\sc MAPPINGS V} code (for the current version of the code, see \citet{MVcode}, also: \citet{Dopita1976, Binette1985, Sutherland1993, Sutherland17}) in a post-processing step\footnote{https://github.com/kativmak/CESS}.
This code assumes CIE, meaning that the ionisation fractions of each element depend only on the gas temperature, with no dependence on the gas density.
This is applicable for SNRs, where the most significant contribution to the (collisionally excited line) emission is produced by shocks and not by photoionisation \citep{Levenson1995, Negus2021}.
We can use the CIE approximation since we are mainly interested in the optical emission of the SNR, i.e. in a fairly evolved SNR, which has departed from the Sedov-Taylor phase.
At the early stages of evolution (when hard X-ray radiation predominates), this approximation should probably not be used, since not all plasma is evenly ionised.
After about $3\times 10^4$~years (for typical densities of 1~cm$^{-3}$), the ionisation state of the plasma no longer plays a role and the gas reaches equilibrium as claimed in \citet{reynolds17_book}. In addition, no significant differences have been found when treating an evolved SNR in equilibrium or in non-equilibrium \citep{Sarkar2021}.

Using {\sc MAPPINGS V} and assuming the WIM metallicity model, we produce a detailed cooling curve as a function of the gas temperature by considering emission from hydrogen, helium, and metal atoms and ions, as well as non-atomic emission such as free-free emission. 
The resulting total emissivity $\Lambda_{\rm N}$ ([erg s$^{-1}$ cm$^{3}$]) is the cooling rate $\Lambda$ ([erg s$^{-1}$ cm$^{-3}$]) divided by $n_{\rm e} n_{\rm H}$, where $n_{\rm e}$ is the electron number density, and $n_{\rm H}$ is the number density of hydrogen nuclei.
We use $n_{\rm e} \approx n_{\rm H} = 1\;{\rm cm}^{-3}$. We show $\Lambda_{\rm N}$ in Fig.~\ref{fig:map_cc} (solid black line).

By modifying the {\sc MAPPINGS V} code, we further divide the total cooling rate into three energy bands ($E_{\rm FUV}, E_{\rm EUV}, E_{\rm X}$) but we could define any other number or limits of energy bands.
We overplot these three emissivities $\Lambda_{\rm N,FUV}$ (dot-dashed grey line), $\Lambda_{\rm N,EUV}$ (dashed brown line), and $\Lambda_{\rm N,X}$ (yellow line) in Fig.~\ref{fig:map_cc}.
Additionally, we show the contributions of the optical cooling lines (dotted lines: S$^+$ in orange; N$^+$ in blue; O$^{++}$ in green; H$_{\rm \beta}$ in cyan; H$_{\rm \alpha}$ in red; and the combination of all hydrogen lines in light grey).
The hydrogen lines dominate at $T\sim 20,000$~K, while N$^+$ and O$^{++}$ are tracing higher temperatures.
In particular, O$^{++}$ is tracing the hotter gas inside the remnant bubble, whereas the other lines (N$^+$, S$^+$ ) are more prominent at colder temperatures, which we expect to find closer to the bubble edge.

\section{Post-processing of SN remnants}
\label{sec:post-processing}

\begin{figure*}
    \centering
    \begin{minipage}[h]{0.49\linewidth}
    \center{\includegraphics[trim={0.5cm 0.3cm 0.3cm 0cm},clip, width=0.7\linewidth]{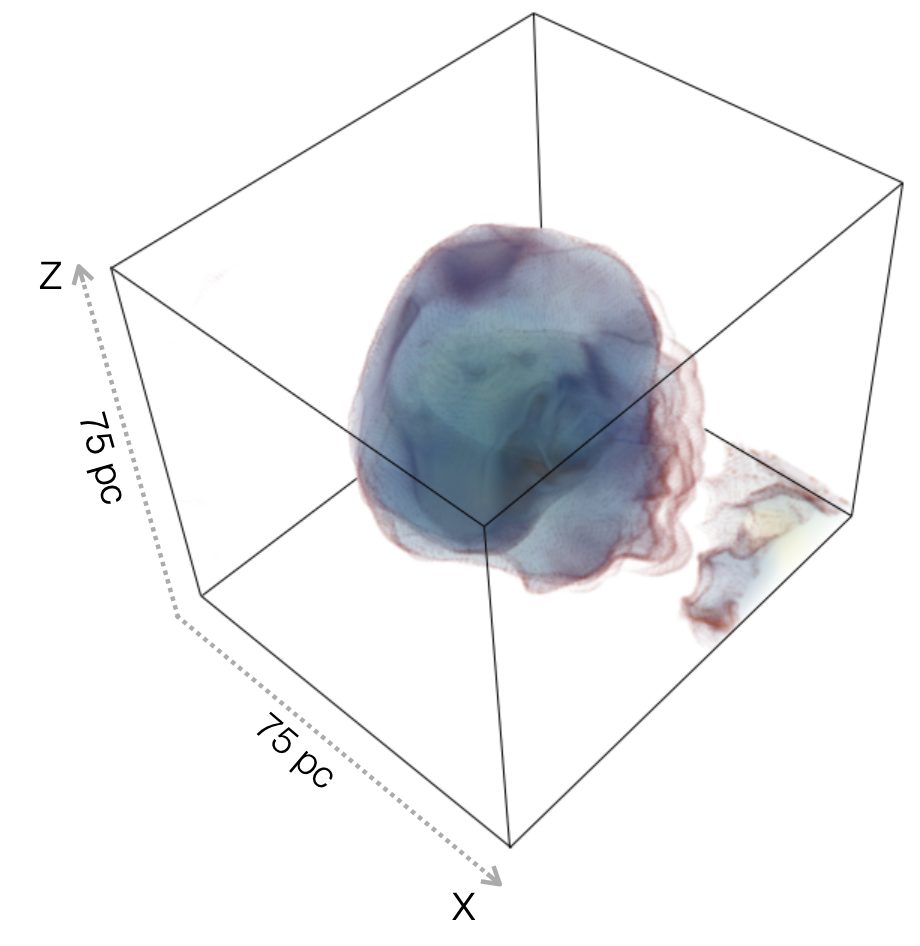} }
    \end{minipage}  
    \hfill
    \begin{minipage}[h]{0.5\linewidth}
    \center{\includegraphics[width=1\linewidth]{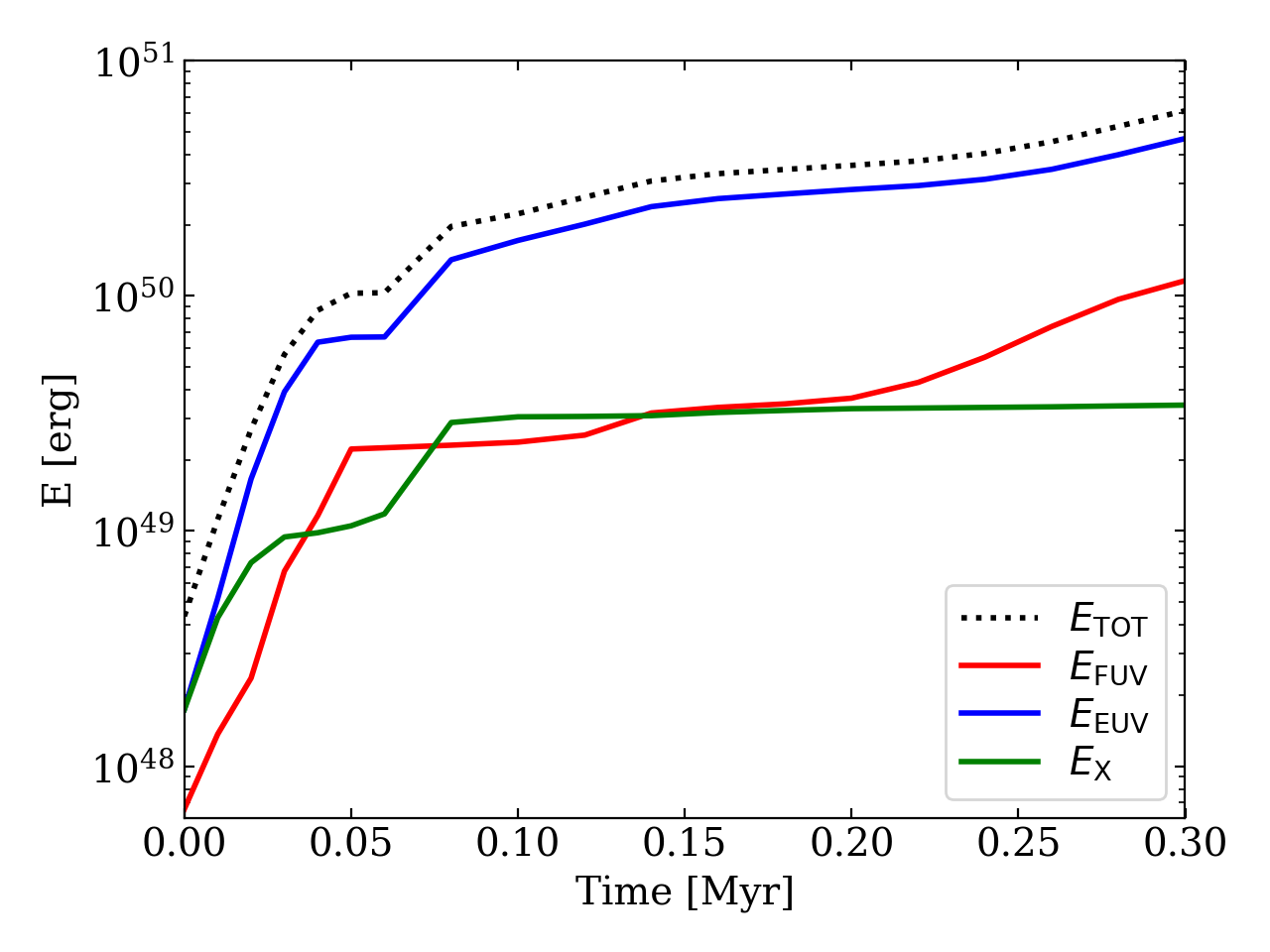}}
    \end{minipage}
    \caption{Left panel: Volume rendering of the post-processed SNR at time $t_{\mathrm{evol}} =0.16$~Myr, showing the FUV/Optical ($E_{\rm FUV} < 13.6$~eV, red), EUV (13.6~eV~$\le E_{\rm EUV} < 100$~eV, blue) and X-ray emission ($E_{\rm X}\ge$~100~eV, green) of the cooling remnant.
    Each side of the cube is approximately 75 pc.
    Note that this is only a part of the simulation box.
    In the lower right corner of the cube, emission from the background is also visible.
    Right panel: Cumulative cooling energy as a function of the remnant's evolution time. We show the total cooled energy (black) and how it is distributed over the three energy bands. The SN explosion energy is $E_\mathrm{SN}=10^{51}$~erg. We find that only a fraction of $E_\mathrm{SN}$ has been radiated away by $t_\mathrm{evol,fin}$. Due to the broad density distribution of the surrounding ISM, $t_\mathrm{TR}$ and $t_\mathrm{PDS}$ vary substantially over the surface of the expanding remnant, leading to an extended period of cooling (see Section~\ref{sec:SNR_times}).}
    \label{fig:energy_3Dsim}
\end{figure*}

\subsection{Time evolution of the cooling energy} \label{sec:cool_energy}
In Fig.~\ref{fig:energy_3Dsim} (left panel) we depict the 3D distribution of the emissivities $\Lambda_{\rm N,FUV}$ (red), $\Lambda_{\rm N,EUV}$ (blue), and $\Lambda_{\rm N,X}$ (green) showing the SNR at $t_{\rm evol}=0.13$~Myr. 
We obtain the local $\Lambda_{\rm N}$ for each cell by linearly interpolating the pre-computed {\sc MAPPINGS V} cooling tables for the given gas temperature of the 3D {\sc Flash} simulation.
Fig.~\ref{fig:energy_3Dsim} only shows a small part of the simulation domain with a side length of about 75~pc in each direction, centred around the SNR.
In the bottom right corner, one can see an emission that is not associated with the young remnant but originates from an older nearby remnant (i.e., background emission).
According to an estimate of the typical star formation rate in a galactic environment with an overall gas surface density of $\Sigma_{\rm gas}=10\;{\rm M}_{\odot}\;{\rm pc}^{-2}$ \citep{Kennicutt1998}, and assuming a standard stellar initial mass function, the typical SN rate in the simulated total volume is $\sim 15$~SN/Myr \citep{Walch15}.
With a typical size of $80-100$~pc \citep{deAvillezBreitschwerdt2007}, old SN remnants are likely to contribute to the emission of a selected sub-region.
We will discuss how to subtract this "background emission" in Section~\ref{sec:SNR_bg} and how the background emission might contaminate the diagnostic of unresolved (not in our Galaxy) SNRs in Section~\ref{sec:BPT}.

On the right panel of Fig.~\ref{fig:energy_3Dsim}, we show the cumulative cooling energy lost by the young SNR as a function of time.
To avoid including the background emission, we explicitly choose a coordinate system within which the explosion centre is located at the origin of the SN bubble and compute the size of the remnant along the principal axes (six directions) for each time step.
To find the edge of the hot bubble, we then search for a sharp drop in the temperature data along each axis.
All $n_{\rm cells}$ cells within the sub-cube defined by the extent of the remnant are used to calculate the total, time-dependent, cumulative cooling energy, $E_{\rm tot}$, as 
\begin{equation} \label{eq:energy}
   E_{\rm tot}(t_{\rm evol})= E_{\rm tot}(t_{\rm evol}-\delta t)+ \delta t \sum_{i=1}^{n_{\rm cells}} \Lambda_{\rm N,i} n_{\rm H,i}^2 V_i,
\end{equation}
where $t_{\rm evol}$ is the current time (see Section~\ref{sec:sim}), $\delta t$ is the time difference between the current time and the previous output. We sum over each cell $i$ with volume $V_i$ in the selected sub-cube. 
Similarly, we compute the cumulative cooling energy in the three different energy bands. 

During the first steps of the SNR evolution, most of the cooling energy is emitted in the EUV and X-ray bands.
When the supersonic blast just hits the surrounding ISM, it heats the gas to X-ray emitting temperatures beyond a few 10$^5$~K (see Fig.~\ref{fig:map_cc}).
As the primary shock wave decelerates, the reverse shock becomes non-radiative and the SNR emits thermal X-ray (and strong radio synchrotron emission).
The physical processes included are free-free radiation (from forward and reverse shocks), inverse-Compton scattering, and X-ray line emission from the ionised metals (Fe, C, O, Ne, etc.).
As can be seen from the right panel in Fig.~\ref{fig:energy_3Dsim}, these first stages do not last long, and the X-ray emission begins to weaken (approximately around 0.01 Myr) as the reverse shock wave has disappeared and the remnant passes into the Sedov-Taylor stage.
Now the energy is mostly released in the EUV, FUV, and optical bands (note that all further low-energy emission, such as infrared emission, is included in this "Optical" band), mainly in the low ionisation stages of elements like sulphur, nitrogen, and oxygen (see Fig.~\ref{fig:energy_3Dsim}) as well as in hydrogen lines.
Since an SNR spends most of its later life radiating in the EUV and optical (before it merges with the surrounding ISM), this emission is a valuable tool for the classification of SNRs and the determination of various parameters such as electron density and temperature of the gas, shock velocity and energy of the SN explosion.
At the time when we stop the simulation ($t_{\rm evol} =0.3$~Myr), the remnant has lost about 60\% of $E_{\rm SN}$. Within the ST phase, it is expected that slightly more than 70\% of $E_{\rm SN}$ are in the form of thermal energy \citep[e.g.][]{Walch&Naab2015, Haid16}. Hence, the remnant has mostly, but not fully cooled down by the time we stop the simulation.

\subsection{Characteristic times for the SNR evolution} \label{sec:SNR_times}
\begin{figure*}
	\includegraphics[width=15cm]{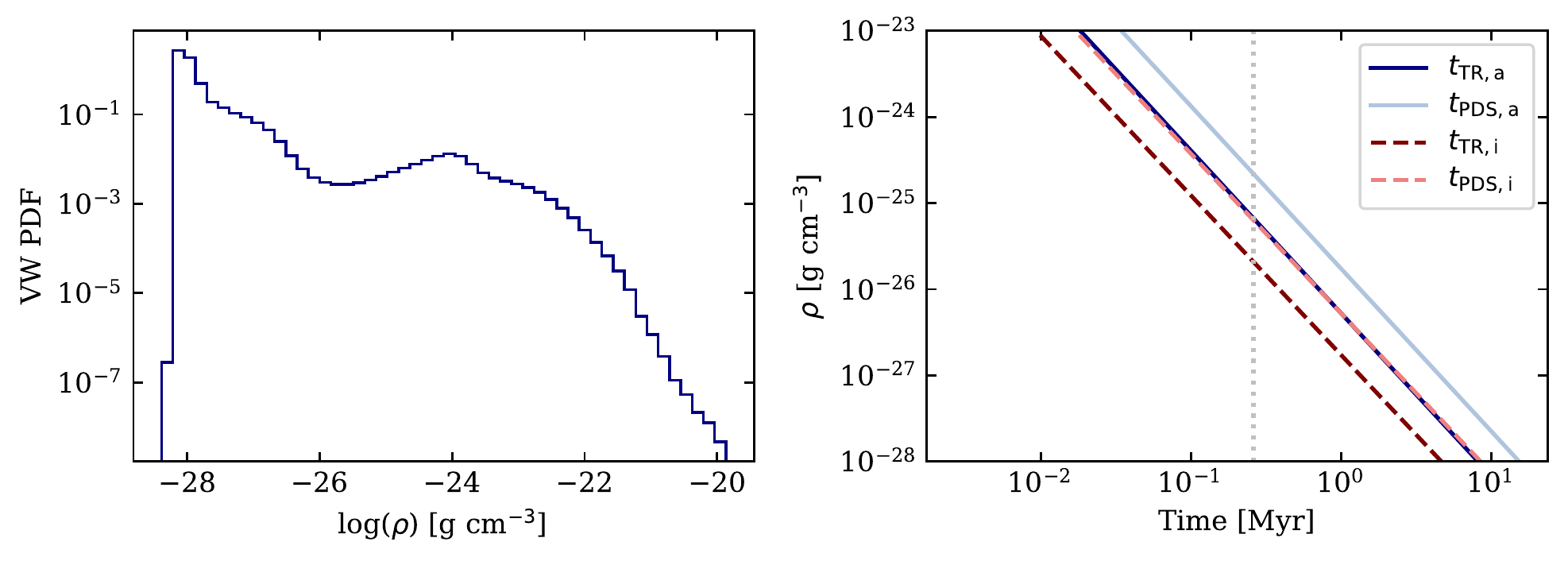}
    \caption{Left panel: volume-weighted density PDF of the initial whole simulation box (t$_{\rm evol}$ = 0.0 Myr). Right panel: Model predictions from \protect\citet{Haid16} for the t$_{\rm TR}$ and the t$_{\rm PDS}$ in ambient atomic media (solid lines) and in ambient ionised media (dashed lines) with different densities. The vertical grey line illustrates the time until we calculate the SNR evolution, t$_{\rm evol}$ = 0.3 Myr.}
    \label{fig:pdf_sn_times}
\end{figure*}
Radiative cooling plays an important role in the evolution of an SNR, especially in an inhomogeneous medium (in our case, next to a MC).
This has been shown in the simple one-dimensional models \citep{Chevalier77, Cioffi88} and then in more complex 3D simulations \citep{Kim2015, Walch&Naab2015}. 
Usually, young SNRs are observed during the ST or TR phase and old SNRs during the radiative phase (TR or PDS). These are typically used characteristic times to describe the SNR evolution. It is important to estimate these times to 
compare synthetic and real observations.
In the ST phase \citep{Sedov59} the remnant only cools adiabatically, while radiative losses are small and can be neglected.
Radiative cooling becomes important in the TR phase. Therefore, the transition to the TR phase is associated with a rise of cooling energy released in the form of optical emission lines due to the temperature dropping below 10$^{6}$ K. 
The PDS phase, in which radiative cooling plays a substantial role, is the longest stage of the SNR evolution before it merges with the ISM.

Due to the complex ISM structure into which the SNR expands in our simulation, it is not possible to calculate unique characteristic times for our remnant.
We demonstrate this in Fig.~\ref{fig:pdf_sn_times}, where we first show the volume-weighted density probability distribution function (PDF) of the surrounding gas at the time when the SN explodes (left panel).
In Appendix A, Fig.~\ref{fig:vw_pdf_evol}, we show the density PDFs for the respective sub-cubes (see Table~\ref{tab:boxes} in Section~\ref{sec:cool_energy} for how the sub-cubes were defined) which contain the SNR at three different times, $t_{\rm evol} = 0.01, 0.13, 0.3$~Myr.
Although most of the volume near the remnant centre is initially filled with gas with a density of $\rho\sim 10^{-23.5}\,{\rm g\, cm}^{-3}$, the overall range of densities is very broad (spanning more than 7 orders of magnitude in density). 
Following the calculation from \citet{Haid16}, we plot the corresponding characteristic evolution times for a given environmental density on the right panel of Fig.~\ref{fig:pdf_sn_times}.
The end time of our simulation is indicated by the vertical grey dashed line. 
We see that the SNR should have evolved well into the PDS stage if it would have exploded in a uniform density medium with $\rho= 10^{-23.5}\,{\rm g\, cm}^{-3}$.
However, expanding into different directions from the point of view of the SNR centre, the characteristic times may differ dramatically.
This is why we cannot state one $t_{\rm TR}$ and one $t_{\rm PDS}$ for the whole SNR: in a simple picture, different parts of the remnant evolve on different characteristic time scales. 

This explains why the remnant is still cooling at $t_{\rm evol}=0.3$~Myr.
The energy evolution of the SNR can be explained as follows: we first have a sharp rise of EUV optical/FUV emission around 0.05 Myr on the right panel of Fig.~\ref{fig:energy_3Dsim}, because it roughly corresponds to $t_{\rm TR,a}$ for the median SN ambient density $\rho \sim 10^{-23.5}$~g~cm$^{-3}$. 
Then the reverse shock travels into the SNR bubble and we can observe the second sharp rise in energy around 0.1 Myr in the X-ray and EUV, but none of the gas is available yet to emit in the optical band (dense gas has not yet reached t$_{\rm PDS}$ and low-density gas has not yet reached t$_{\rm TR}$). 
There is a part of low-density ISM gas left (around 10$^{-25}$ -- 10$^{-26}$ g cm$^{-3}$) which can finally reach t$_{\rm TR}$ and emit in the optical/FUV band significantly later. 
As a result, we do not see a decrease in the optical/FUV emission after 0.2 Myr of SNR evolution. Only at 0.25 Myr dense gas will be at the PDS stage and part of the low-density gas at the transition phase. No doubt that we need a longer timescale for all gas to reach a PDS stage. 
However, 0.3 Myr is still a long enough timescale to cover all stages of SNR evolution for the most part of the SNR bubble.

\subsection{SNR interaction with a MC}
\begin{figure*}
	\includegraphics[trim={0.0cm 0.0cm 0.0cm 0.0cm},clip, width=14cm]{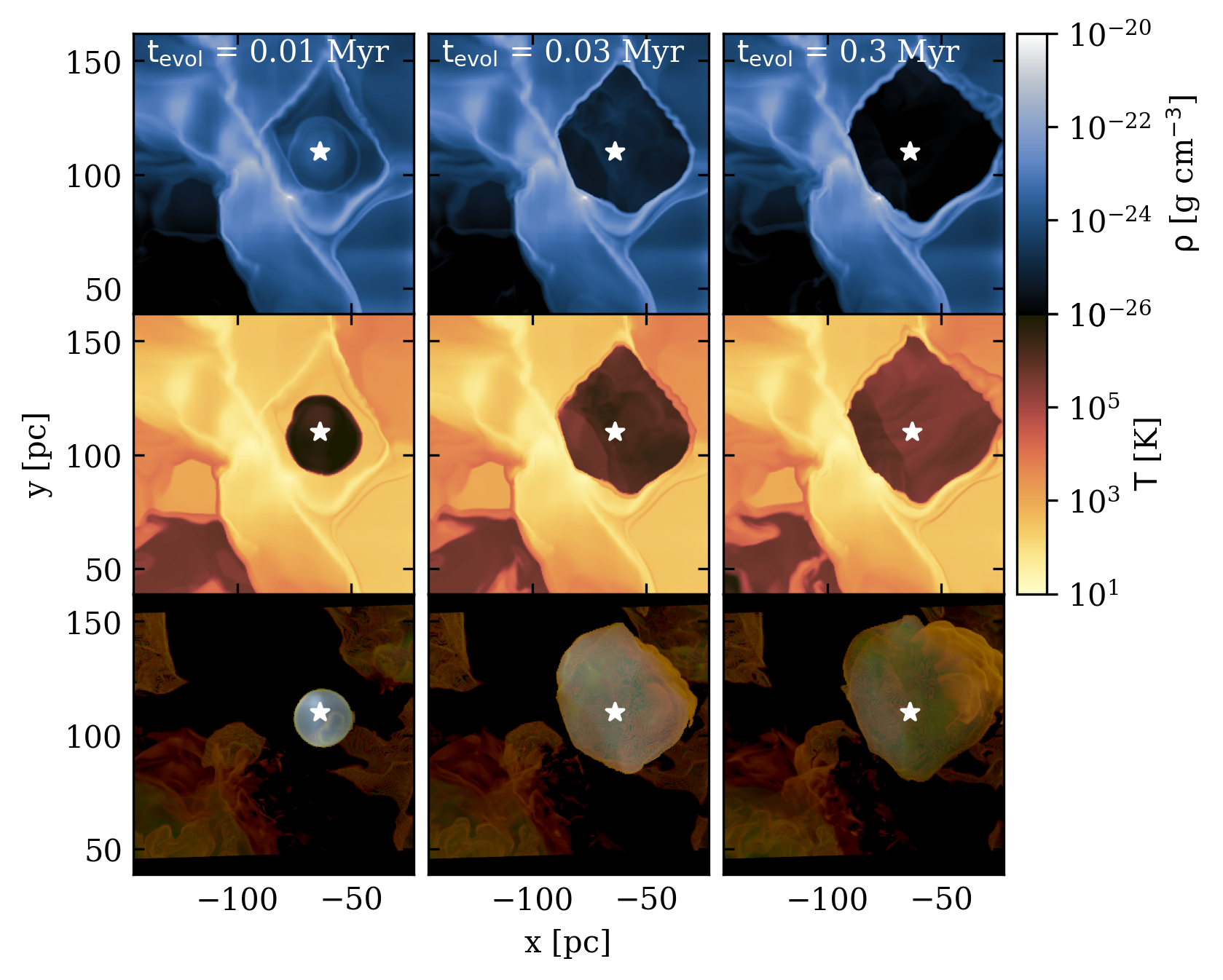}
    \caption{Time evolution (from left to right) of the density (top row) and temperature (middle row) in slices through the explosion centre.
    The bottom row shows the RGB projection image of the emitted energies in $E_{\mathrm{FUV}},\; E_{\mathrm{ EUV}},\;{\mathrm{and}}\; E_{\mathrm{X}}$.
    The SN position is shown as a white star symbol.
    The radiation in the upper left corner is not a result of this SN explosion: it is a leftover from other older SNe from the background, see details in \protect\citet{Seifried18}.
    The dense MC is located to the left of the explosion centre.
    The distance between the explosion centre and the cloud's centre of mass is 25~pc.
    The remnant expands in a highly asymmetric fashion as the SN shock dissipates quickly when it encounters the MC.}
    \label{fig:t_d_bands}
\end{figure*}

This work will consider emission before, during, and after the TR stage for the densest part of the SNR bubble for comparison (t$_{\rm evol}$ = 0.01, 0.13, 0.3 Myr).
In Fig.~\ref{fig:t_d_bands}, from top to bottom, we show the time evolution (from left to right) of the density and temperature in slices through the explosion centre and RGB images of the three different energy bands (X-ray in blue, EUV in green, optical/FUV in red).
The SNR bubble is heated to more than $10^6$~K, leading to the hydrodynamic expansion of the remnant.
Due to the turbulent surrounding medium, the expansion of the remnant is highly asymmetric. 

Fig.~\ref{fig:t_d_bands} clearly shows that even though the SN explosion is initially symmetric, the MC distribution dominates in shaping the remnant morphology \citep{Kafatos1980}.
The density of the MC gas is several orders of magnitude higher than in the low-density ISM ($n_{\rm H} > 10^3 \: \mathrm{cm}^{-3}$).
Therefore, the velocity of the shock wave towards the centre of the cloud is much smaller and, in some places, the shock almost stops not able to propagate further into the MC.
Hitting a dense medium also leads to rapid gas ionisation.
It launches a strong reverse shock that travels into the ejecta.
Therefore, the hot interior of the SNR is a source of X-ray emission like we see in observations \citep[see][or the overview by \citealt{Vink2012})] {Slane1999, Long1991}\footnote{This can be seen in the attached animation, in particular in the X-ray band.}.
Moreover, the explosion cannot spread to the centre of the MC (the area of highest density to the left of the explosion centre), but it continues to expand into the more diffuse region towards the upper right corner of the emission maps, forming dense filaments as a result of instabilities.
These structures are an essential morphological feature as a criterion for the interaction of an SNR with a cloud, which is also commonly found in real observations \citep[see][and others]{Fesen1997, Boumis2002, Boumis2004, Katsuda2016,How2018}.
In general, we can conclude that the SNR has a very complex structure.
In projection, the shocked filaments which are visible in the Optical and EUV overlap with the volume-filling X-ray emission (see also Fig.~\ref{fig:energy_3Dsim} for the 3D example).\\

\begin{table}
    \centering
    \begin{tabular}{c|c|c|c|c|c|c}
        \hline
        t$_{\rm evol}$ & a$_{\rm x,SNbox}$ & a$_{\rm y,SNbox}$ & a$_{\rm z,SNbox}$ &s$_x$ & s$_y$ & s$_z$ \\
        Myr & [pc] & [pc] & [pc] & [pc] & [pc] & [pc] \\
        \hline
        0.01 & 64.8 & 77.5 & 68.1 & 16.2 & 19.5 & 16.1\\
        0.13 & 64.8 & 77.5 & 68.1 & 32.4 & 34.4 & 40.1 \\
        0.3 & 64.8 & 77.5 & 68.1 & 56 & 66 & 55.1 \\
        \hline
    \end{tabular}
    \caption{Time evolution (column 1) of the SN box ($a_{x,y,z}$, columns 2-4) and a shock box (defined from our shock detection routine) along the x,y and z-axis ($s_{x,y,z}$, columns 5-7). We use shock boxes to calculate the parameters (temperature, density and luminosity) inside and outside of the SN bubble at different times for Fig.~\ref{fig:2phase_diagram}. SN box is used in all other cases.}
    \label{tab:boxes}
\end{table}

\begin{figure*}
    \centering
        \includegraphics[trim={0cm 0.2cm 0.0cm 0.2cm},clip, width=0.68\textwidth]{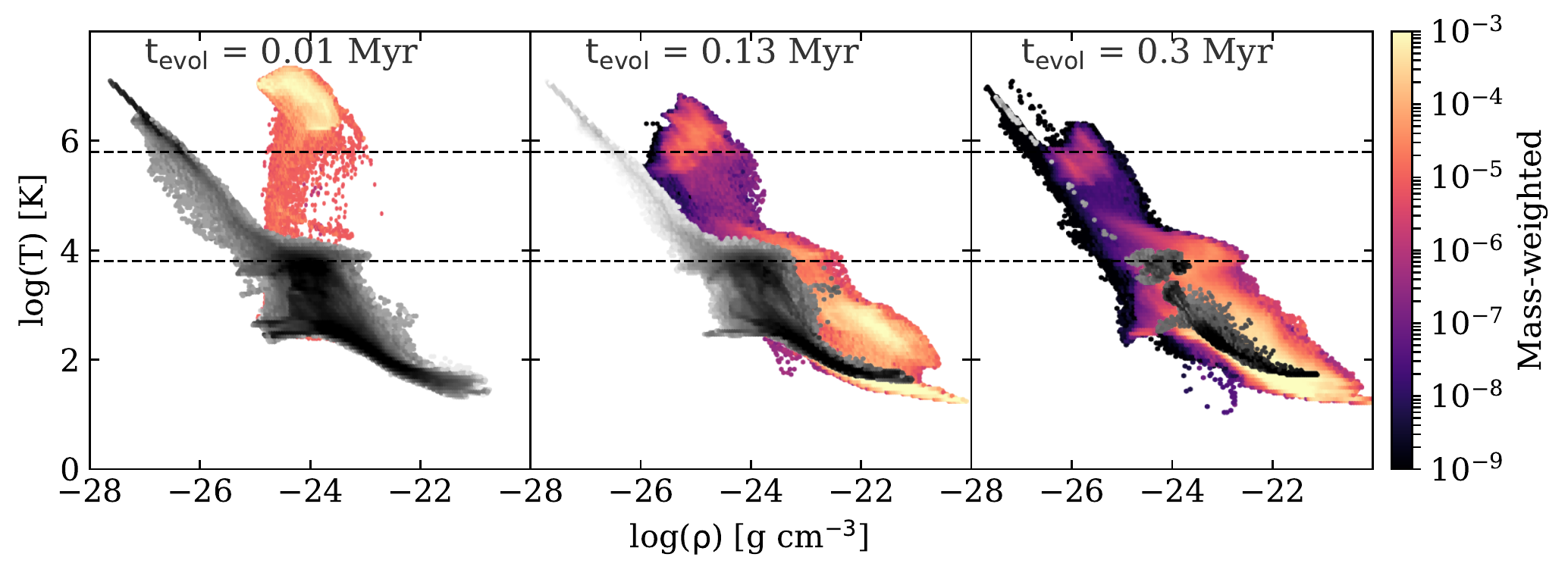}
        \includegraphics[trim={0cm 2.3cm 0.0cm 0.3cm},clip, width=0.7\textwidth]{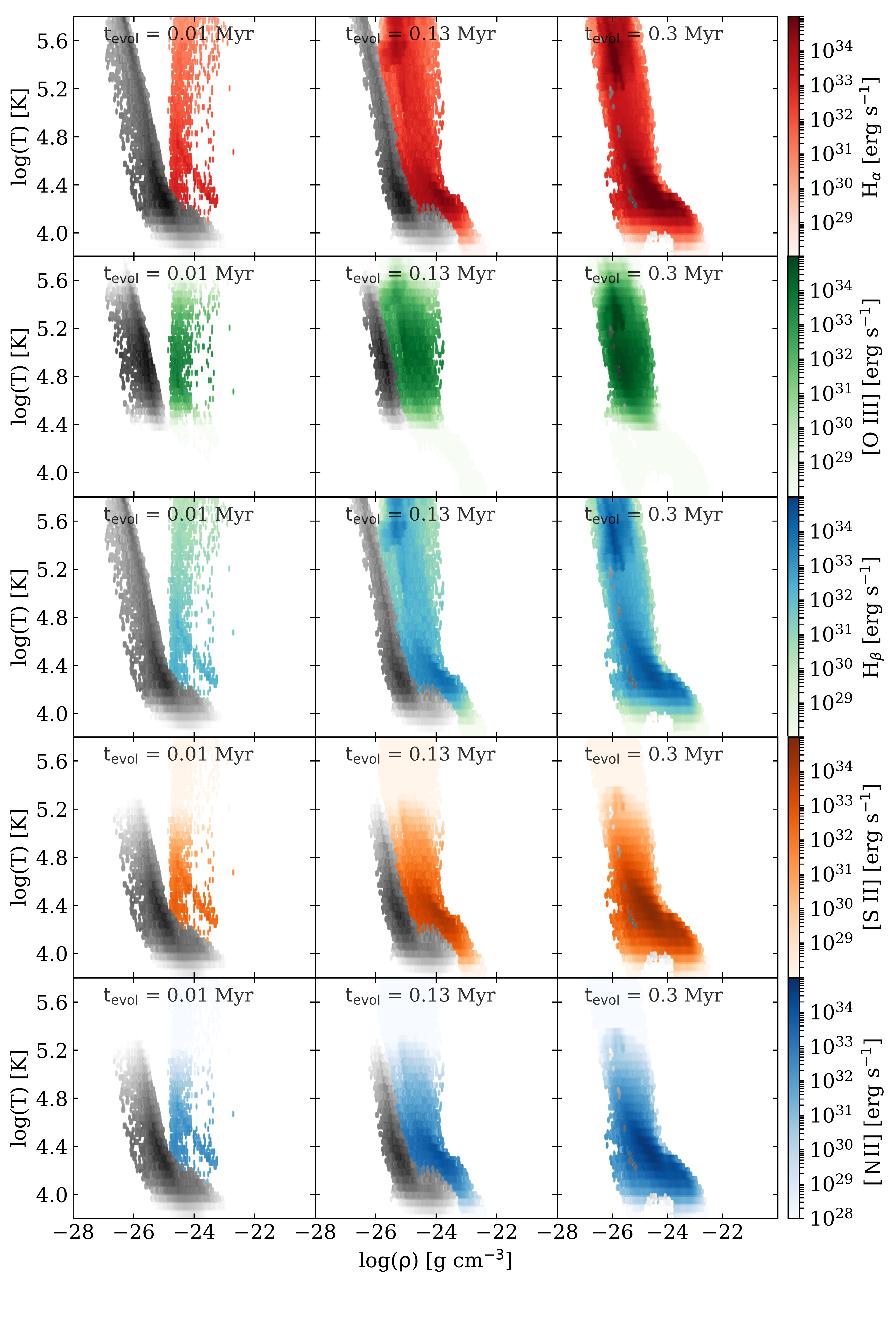}

    \caption{\textbf{The top row}: Mass-weighted 2D PDFs showing the gas distribution in the density-temperature phase space for three different times (from left to right). "Coloured" is all gas from within, or nearby, the SNR bubble, and "grey" is what we identify as background gas that is unaffected by the SNR.
    The dashed horizontal lines mark the temperature range that is selected for the emission-weighted PDFs shown below.
    \textbf{2-5 rows from the top:} gas distributions in the temperature-density phase space colour-coded with the calculated optical emission lines (from top to bottom: H$_{\alpha}$, [O III] ($\lambda 5007$), H$_{\beta}$, [SII] ($\lambda 6717$, $\lambda 6731$), and [N II] ($\lambda 6583$)). Again, the background gas is shown in grey-scale.
    The gas distribution at time $\mathrm{t_{evol} = 0.1}$ Myr has a "V-shape": the right branch is produced by the current SN explosion while the left branch is the background material heated by the previous SNe.
    As the current SN evolves, the two branches join because the bubble expands and the gas cools.}
    \label{fig:2phase_diagram}
\end{figure*}

From our 3D simulation data, we plot mass-weighted 2D PDFs in Fig.~\ref{fig:2phase_diagram}, which show the distribution of the multi-phase ISM gas in the temperature-density phase space (top row).
From left to right, we show the same times as in Fig.~\ref{fig:t_d_bands}, i.e. $t_{\rm evol}= 0.01,\;0.13,\;{\rm and}\; 0.3$~Myr after the explosion took place. 
Cells which are identified to belong to the SNR sub-cube are shown in colour, while all other cells within the SILCC zoom-in region are shown in grey-scale as these are considered to belong to the background.
Due to the highly asymmetric morphology of the remnant, the sub-cube also contains some gas from the shell and the nearby cloud. However, it is quite obvious which gas belongs to the hot interior and the cooling shell and which material is part of the surrounding ISM contained in the sub-cube.
At $t_{\rm evol}=0.01$~Myr, the young remnant can be clearly seen as it incorporates relatively dense gas (the gas near the explosion centre initially has number densities between $n \sim 1-10\;{\rm cm}^{-3}$), which is heated to $T\gtrsim 10^{6-7}$~K at $t_{\rm evol}= 0.01$~Myr. The thin shell around the SNR bubble is also distinguishable at this time step as an almost horizontal branch at $T\gtrsim 10^{4.3}$~K.
The hot bubble moves to the left, to lower densities, as the remnant expands, and it almost merges with the hot gas "branch" at lower densities, which is populated by older background SN remnants.

Using {\sc MAPPINGS V}, we calculate the luminosity of key optical emission lines for each grid cell following Eq.~\ref{eq:energy}.
We show the luminosity-weighted 2D PDFs in rows 2-6 of Fig.~\ref{fig:2phase_diagram}. The $y$-axis range is limited to show the warm-hot medium with temperatures between $10^{3.8}~{\rm K}< T < 10^{5.8}~{\rm K}$ (as indicated by the dashed horizontal lines in the top panels), as this is the temperature regime where the optical lines are bright (and can physically be exited, see Fig.~\ref{fig:map_cc}). Again, we distinguish between material within the SN sub-cube (coloured distribution) and background gas (grey-scale). 
We plot H$_{\alpha}$ (2$^{\rm nd}$ row), [O III] ($\lambda 5007$) (3$^{\rm nd}$ row), H$_{\beta}$ (4$^{\rm th}$ row), [S II] ($\lambda 6717, 6731$) (5$^{\rm th}$ row) and [N II] ($\lambda 6583$) (6$^{\rm th}$ row, one can see only one sulphur line in Fig.~\ref{fig:2phase_diagram} because [S II] ($\lambda 6717$) and [S II] ($\lambda 6731$) have the same behaviour and properties). For emission lines in Fig.~\ref{fig:2phase_diagram}, it is impossible to draw a simple luminosity-density dependence. 

\section{Optical emission line diagnostics}
\label{sec:optical_emiss}
\subsection{Background subtraction: resolved vs. unresolved SNRs}
\label{sec:SNR_bg}

The background of our simulation may contain emissions from previous SNRs (for example, see the lower right corner on the left panel of Fig.~ \ref{fig:energy_3Dsim}), which are not the subject of our work but can represent a realistic background as in observations. 

In order to carry out the emission line diagnostics, we include the option to reduce the background of the signal, since we can calculate the emission of the warm-hot, diffuse gas right before the SN explosion takes place. The background is defined as the emission calculated for the gas distribution immediately before the SN explosion (at t$_{\rm evol}$ = 0.0 Myr).
In this way, we compute the background emissivity for each line in 3D. The lines we consider are H$\alpha$ ($\mathrm{\lambda 6563}$), H$\beta$ ($\mathrm{\lambda 4861}$), [O III] ($\mathrm{\lambda 5007}$), [N II]($\mathrm{\lambda 6583}$), [S II] ($\mathrm{\lambda 6717}$), and [S II] ($\mathrm{\lambda 6731}$).
Hence, for every snapshot we analyse, we calculate the line emissivities in 3D and, if background subtraction is switched on, we just subtract the 3D cube of the background from the current 3D cube of cooling emission.

We may now choose a viewing angle, i.e. the position of the outside observer and integrate the 3D cube along the line-of-sight (LOS) to obtain the corresponding flux map.
We integrate over the map to obtain the total luminosity ($L_{\rm tot}$, [erg s$^{-1}$]) of a given line.
Assuming the gas to be optically thin, the resulting total luminosity of the background, $L_{\rm bg}$, is given in Table~\ref{tab:background} for each of the computed lines.
Since the background emission is assumed to be constant in time while the actual line emission from the SNR evolves as a function of time, the background subtraction reduces the overall line flux by different amounts, i.e. the "attenuation percentage" changes.
We list the minimum and maximum values of the attenuation percentage in the third column of Table~\ref{tab:background}.

\begin{table}
    \centering
    \begin{tabular}{|c|c|rc}
        \hline
        Line & $L_{\rm bg}$ & Min \%, Max \%\\
         & [erg s$^{-1}$]\\
        \hline
        H$\alpha$ ($\mathrm{\lambda 6563}$) & 1.01 $\times \, 10^{38}$ & 8.5\% - 48.0\% \\
        H$\beta$ ($\mathrm{\lambda 4861}$) & 2.21 $\times \, 10^{37}$& 7.3\% - 48.0\% \\
        $[$OIII] ($\mathrm{\lambda 5007}$) & 7.93 $\times \, 10^{37}$ & 2.5\% - 35.0\% \\
        $[$NII] ($\mathrm{\lambda 6583}$) & 4.16 $\times \, 10^{37}$ & 10.5\% - 42.5\% \\
        $[$SII] ($\mathrm{\lambda 6717}$) & 2.56 $\times \, 10^{37}$& 17.5\% - 81.0\% \\
        $[$SII] ($\mathrm{\lambda 6731}$) & 1.82 $\times \, 10^{37}$ & 15.5\% - 76.5\% \\
        \hline        
    \end{tabular}
    \caption{The constant integrated background luminosity for each optical emission line (as indicated in column 1) for the optically thin case (column 2).
    A percentage showing how much the background subtraction reduces the overall line flux during the time evolution of SNR (column 3; see section~\ref{sec:SNR_bg}).}
    \label{tab:background}
\end{table}

In the course of the paper, we follow two different approaches, motivated by real observations: in case the SNR would be located in our Galaxy and could be well resolved \citep[pixel-by-pixel resolved SNR, for example,][]{Fesen1985, Boumis2022}, we should apply the background subtraction and study the resolved line intensity maps; if the SNR would be located in some nearby galaxy, it would appear as an unresolved SNR and we only consider the integrated line luminosity. In order to mimic the real observations, we generally do not subtract the background in this case, as an unresolved object could not be disentangled from the diffuse emission which is contained in the same beam.
However, we discuss resolved maps without background subtraction as well as unresolved maps in case a background subtraction would be possible throughout the next sections. In this way, the simulations can be used to inform real observations on this matter.

\begin{figure*}
	\includegraphics[width=16cm]{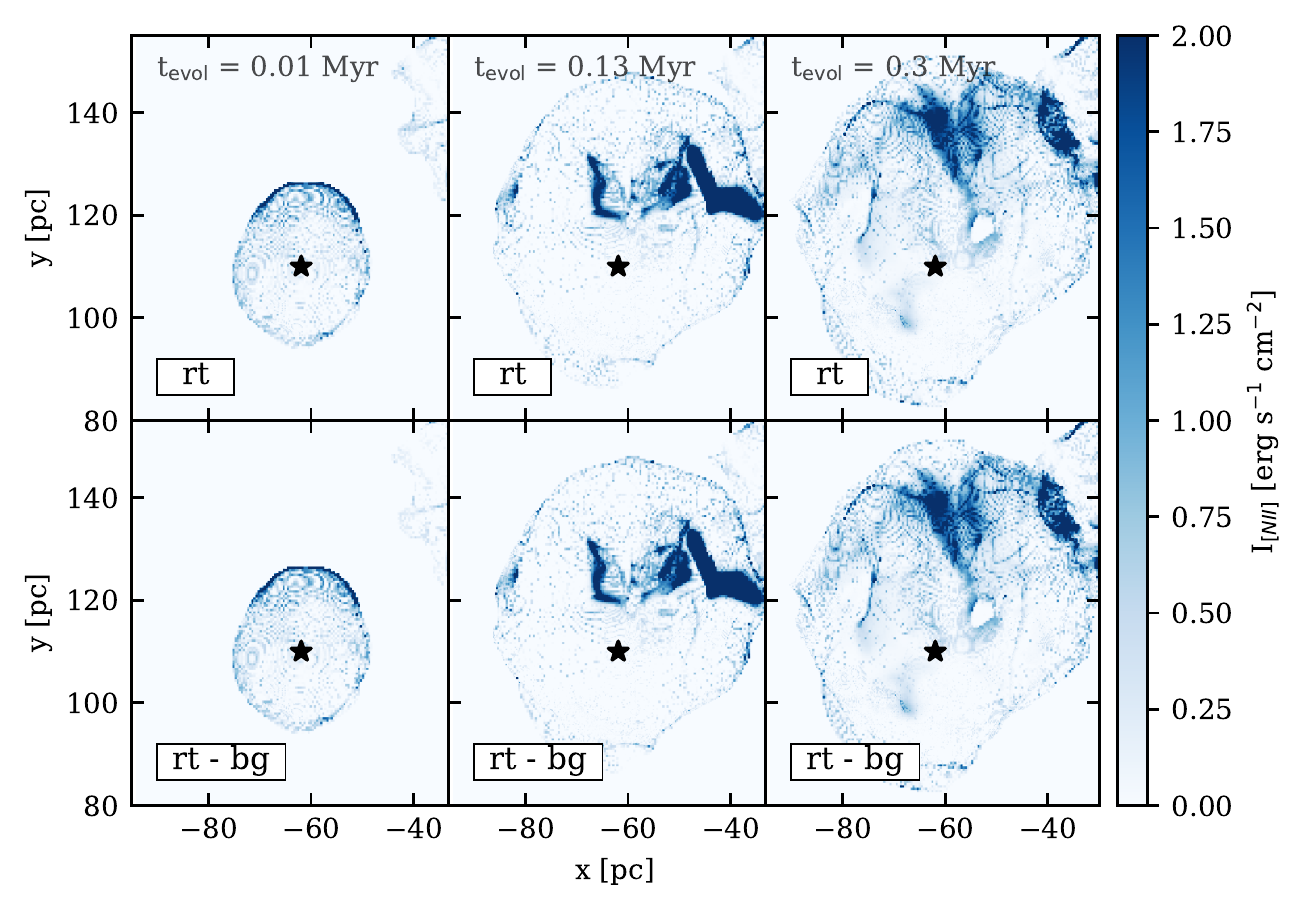}\\
    \caption{Time evolution (from left to right) of the [N II] $\mathrm{\lambda 6583}$ emission: top row with a simple radiative transfer, bottom row with a simple radiative transfer and background subtraction. Both rows show the $x$-axis projection from the top view. The intensity is normalised to the value $10^{-17}$.The intensity of the bubble is the same in both cases, but the background emission is less intense with the background subtraction procedure.}
    \label{fig:n2_row}
\end{figure*}

An example of the optical emission maps is shown in Fig.~\ref{fig:n2_row} and in the Appendix \ref{appendix:emission} for the projection along the $x$-axis from the top ('t') view. 
A typical observational limit used for real telescopes of $3 \times 10^{-18} \: \mathrm{erg \: s^{-1} cm^{-2} arcsec^{-2}}$ \citep[inspired by the Local Volume Mapper (LVM) targeting the Milky Way at $37"\,$ \char`~ $\,$ pc scales, $3\sigma$ sensitivity;][]{LVM, h2_lim2021} is not shown here because the lowest intensity is above this limit. Note that we use a single observational limit for all lines. 

It is apparent, that the different lines trace different environments within the remnant: one originates from the thin layer tracing the recombination zone along the bubble rim and is bright in H$_{\beta}$, [NII], and [SII], while the other one originating from volume-filling gas within the bubble is bright in H$_{\alpha}$ and [OIII].

\subsection{Calculation of line intensity maps}\label{sec:tau}

Typically, when observing SNRs in our Galaxy, one does not need to consider the absorption of the cooling radiation by the surrounding gas because of the low ambient gas densities (see Section~\ref{sec:intro}).
However, the optical line emission can become opaque if the SN explodes in a relatively dense medium or interacts with a MC.
This applies to approximately 10\%-20\% of all SNe in our Galaxy \citep{Hewit&Yusef-Zadeh2009}.
The SNR considered here does interact with a MC.
Hence, we check how much the different optical lines are affected by dust absorption along the LOS.

To get an idea of the amount of absorption, we calculate the optical depth for each cell $i$ in the computational domain as:
\begin{equation}
    \tau_i = \kappa_{\rm abs} \rho_i V_i^{1/3} f_{\rm d},
\end{equation}
where $\kappa_{\rm abs}$ is the dust absorption cross section per mass of dust ($\mathrm{cm^{2} g^{-1}}$), $\rm \rho_i$ is the density of cell $i$, $V_i$ is the cell volume, and $f_{\rm d}$ is the dust-to-gas ratio ($f_{\rm d} = 0.01$ is fixed in our simulations).
The dust absorption cross-section is taken from \citet{WeingartnerDraine2001}: we use the silicate model for Milky Way dust with $\mathrm{R_V} = 4.0$.

For each projection axis (we consider projections along each principal axis, i.e. along the $x-$, $y-$, and $z-$direction), we may choose to start the LOS at the top ('t'; solid lines) or the bottom ('b'; dashed lines), respectively. Then we calculate the integrated flux:
\begin{equation}
    F_{\rm tot} = \int F_i e^{- \tau_{\rm i}} \,ds 
\end{equation}
where $F_{\rm i}$ is the flux of the cell $i$, $\tau_{\rm i}$ is the optical depth, $ds$ is the area of the cube.

\begin{figure}
    \centering
    \includegraphics[trim={0.8cm 0.8cm 0.5cm 0.8cm},clip,width=\linewidth]{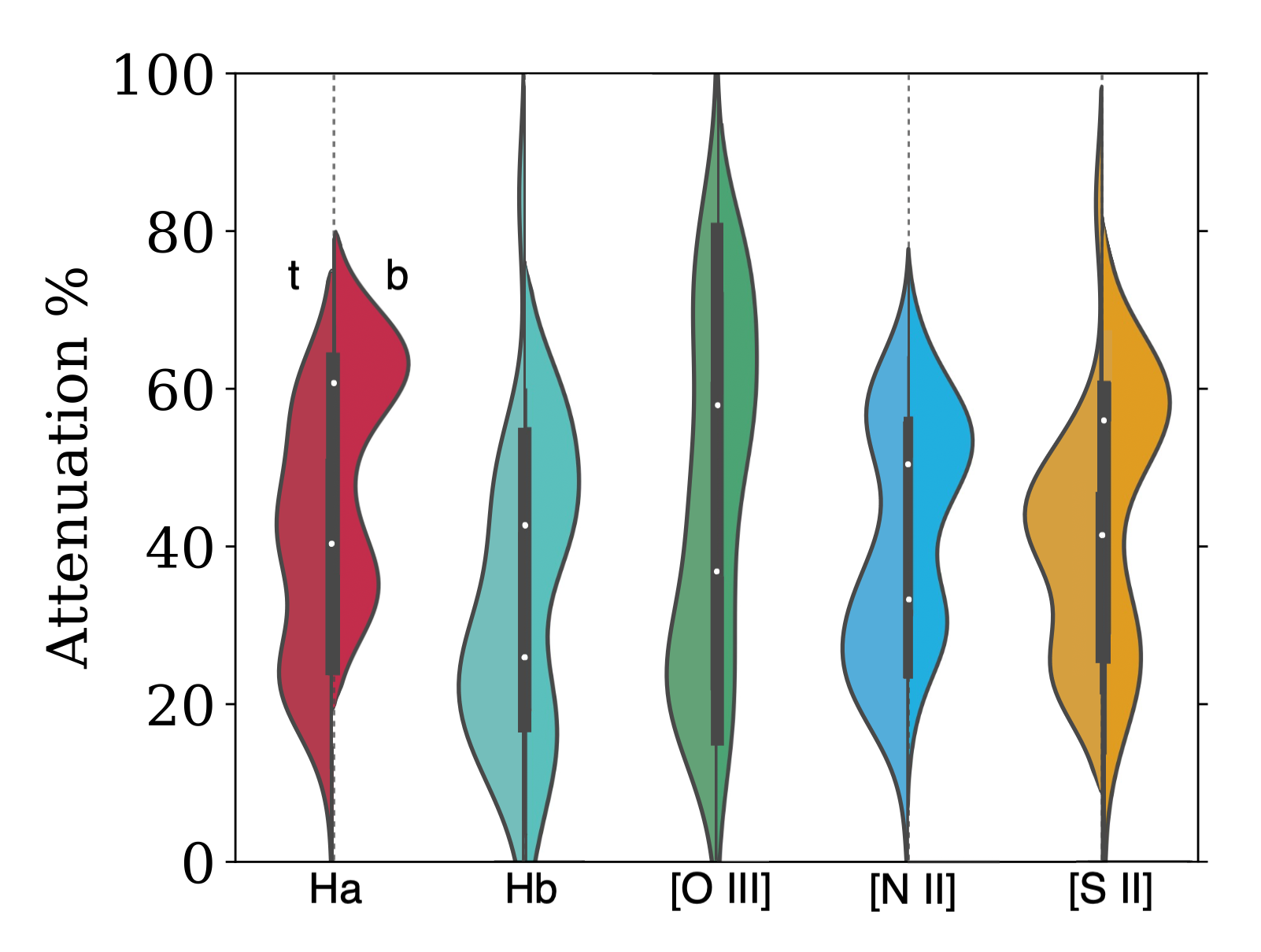}
    \caption{Violin showing the distribution of the percentage of the attenuated flux for 0.3 Myr for each optical line. The white dot is the median for the corresponding value, the tick bar in the centre is the interquartile range.
    The labels top ('t') and bottom ('b') correspond to different viewing angles of a potential observer looking at the cube from the top or the bottom. 't' and 'b' views are separated for every violin plot by the dashed vertical line. The maximum attenuation value varies for every optical line and differs for the 't' and 'b' views.}
    \label{fig:violin}
\end{figure}

In Fig.~\ref{fig:violin}, we show the attenuation percentage for the 3 LOS together for the 'top' view and for the 'bottom view' (see Section~\ref{sec:SNR_bg} and Table~\ref{tab:background}) as during the evolution time for six different emission lines.
The violin plot shows the peak in the data using the kernel density plot. The median values of the 'top' and 'bottom' views are shown with white dots. The higher probability to find a value in the dataset, the thicker the violin plot is. 

The attenuation percentage varies between $\sim$ 5\% - 85\%, depending on the viewing angle.
For example, when looking at the remnant projected along the $x$-axis from the 'top', the dense MC is located between the remnant and the observer.
Hence, the attenuation for this LOS is significant, in particular for later times.
This is not the case when "observing" the cube from the bottom, as there is very little dense gas in front of the remnant.
Although the H$_{\beta}$-line has the shortest wavelength and hence the highest $\kappa_{\mathrm{abs}}$, its attenuation is mostly weaker than for e.g. [OIII].
This indicates that the geometry of the emitting region probably determines the different attenuation percentages of the inspected lines, while the difference in $\kappa_{\mathrm{abs}}$ is less important. 

\begin{figure}
    \centering
    \includegraphics[trim={0.3cm 0.3cm 0.2cm 0.2cm},clip,width=\linewidth]{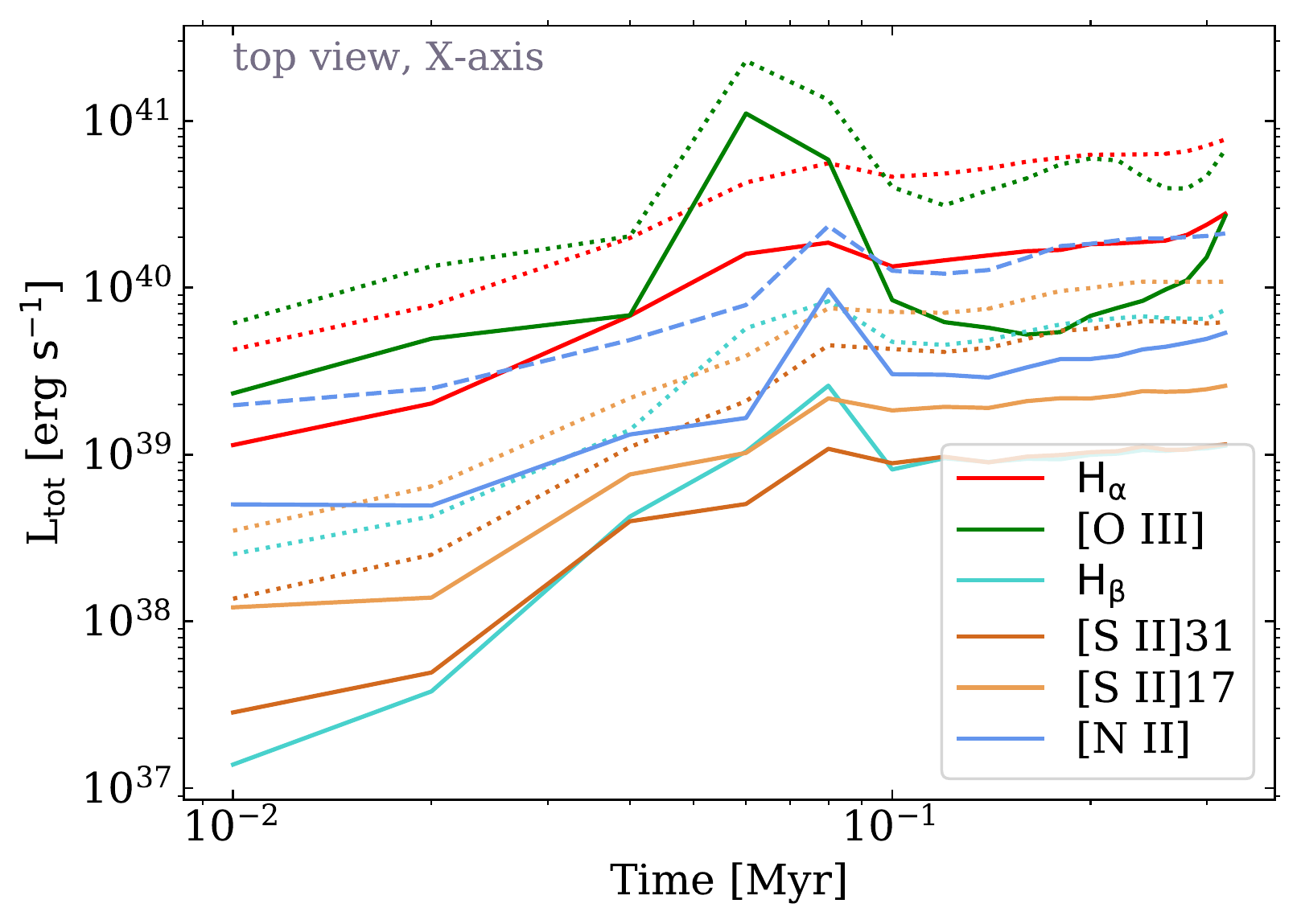}
    \caption{Time evolution of the total luminosity with attenuation (solid lines) and without (dotted lines) for each forbidden optical line. We show only the top view along the X-axis: the top view corresponds to the viewing angle of the potential observer looking at the cube. Other axes show a similar time evolution with the same order of magnitude of L$_{\rm tot}$.}
    \label{fig:flux_att}
\end{figure}

In Fig.~\ref{fig:flux_att}, we plot the time evolution of the total luminosity (i.e. the sum of $F_{\rm tot}$ over all pixels), $L_{\rm tot}$, for the six different lines (H$\alpha$ ($\mathrm{\lambda 6563}$), H$\beta$ ($\mathrm{\lambda 4861}$), [O III] ($\mathrm{\lambda 5007}$), [N II] ($\mathrm{\lambda 6583}$), [S II] ($\mathrm{\lambda 6717}$), and [S II] ($\mathrm{\lambda 6731}$)).
We do not show here the 't' and 'b' views along different axes due to the similar behaviour of $L_{\rm tot}$ values. The solid lines depict the line luminosity which reaches the observer when taking into account the attenuation ('RT'), while the dotted lines show the optically thin case ('no RT').
At a typical distance of $\sim 5.6$~kpc, the LVM detection limit corresponds to a required minimum luminosity of $L_{\rm tot}$ of $3.4 \times 10^{25}\;{\rm erg\; s}^{-1}$ (K. Kreckel, priv. comm.).

At early times, we have only weak radiation in the optical range (consistent with the right panel of Fig.~\ref{fig:energy_3Dsim} showing that all the energy goes to the EUV and X-ray bands).
This is followed by a sharp rise in [O III], which is mainly formed at higher temperatures than [S II] and [N II], namely in the volume-filling bubble interior (see Fig.~\ref{fig:2phase_diagram}).
Shortly after, there is a rise in sulphur and nitrogen, when a recombination layer appears at lower temperatures.

\subsection{Emission line ratios}

\begin{figure*}
	\includegraphics[width=15cm]{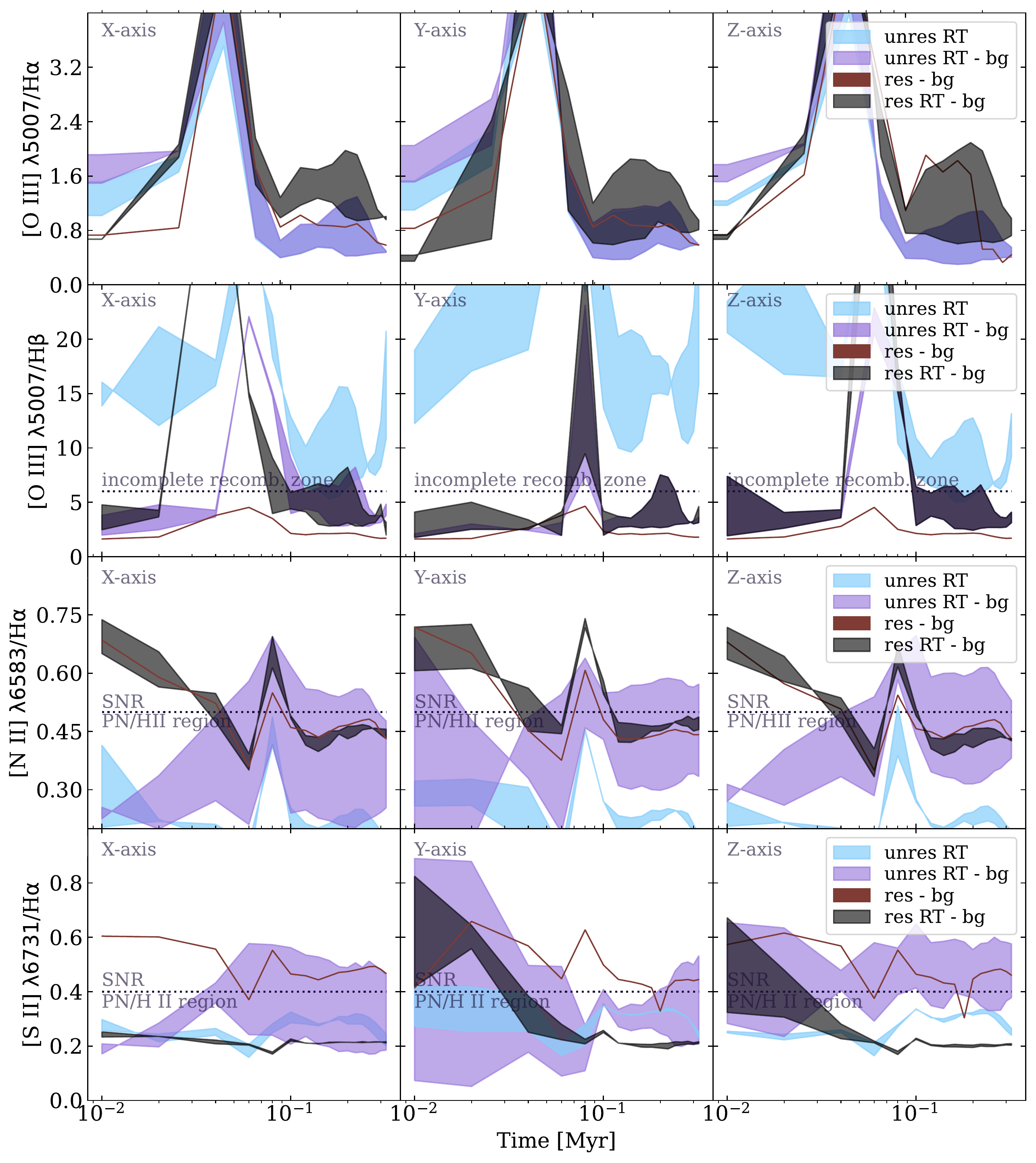}
    \caption{Time evolution of the relevant line ratios for presumably resolved (brown and grey) and unresolved (blue and violet) SNRs for three different projections (left to right). For the [S II] $\mathrm{\lambda 6731}$/H$\mathrm{\alpha}$ ratio in unresolved SNRs we use the sum of the two close sulphur lines ([S II] $\mathrm{\lambda 6717 + 6731}$) because they are in general not resolvable when the SNR is unresolved). For curves labelled with '-bg' the background has been subtracted; all curves labelled with 'RT' consider the line optical depth (see Section\ref{sec:tau}). The shaded regions show partly substantial variations in the line ratios resulting from the two considered observer positions (top and bottom views). These are equal for the presumably optically thin case ('res -bg'). 
   The dotted line shows the theoretical transition between different types of objects (SNR, PN, or HII region).
    While [OIII] is boosted with respect to H$_{\rm \alpha}$ and H$_{\rm \beta}$, the sulphur line ratio is decreased by optical depth effects. 
    We find that a unique classification as an SNR is not possible at all times. Background subtraction plays a key role in all line ratios.}
    \label{fig:line_ratios}
\end{figure*}

Usually, SNe exploding in a dense environment reveal their presence in the form of narrow emission lines in their optical spectra.
These collisionally excited lines play an important role in classifying an object as an SNR and determining its physical parameters such as electron temperature, $T_e$, or electron density, $n_e$.
Once the emission lines and their ratios have been obtained, the following steps can classify an object as an SNR (in the resolved case) according to \citet{Mathewson1972, Dopita84, Blair85, Fesen1985, Kewley2001, Ciardullo2002}:

\begin{enumerate}
  \item $[$O III$]$ $\mathrm{\lambda 5007}$/H$\mathrm{\alpha}$ (sensitive to $T_{\rm e}/n_{ \rm e}$), 
  \item $[$O III$]$ $\mathrm{\lambda 5007}$/H$\mathrm{\beta}$ (completeness of recombination layer; see Section\ref{sec:BPT}),
  \item  $[$N II$]$ $\mathrm{\lambda 6583}$/H$\mathrm{\alpha} > 0.5$,
  \item  $[$S II$]$ $\mathrm{\lambda 6731}$/H$\mathrm{\alpha} > 0.4$.
\end{enumerate}

The given line ratio thresholds were first derived based on multiple observations of SNe in our Galaxy, then derived in a purely theoretical optical classification scheme by \citet{Kewley2001} using a combination of stellar population synthesis, photoionisation, and shock models, and then improved significantly by observational data from the SDSS survey by \citet{Kauffmann2003}.

The time evolution of these optical line ratios is shown in Fig.~\ref{fig:line_ratios} (with the actual maximum and minimum values summarised in Tables \ref{tab:line_ratios_res} and \ref{tab:line_ratios_unres} in Appendix~\ref{appendix:emission}). 
Due to the uneven evolution of the SNR bubble (as was described in Section~\ref{sec:SNR_times}), our object cannot be classified as an SNR for every time step, but the line ratios mainly follow the values that are commonly observed for SNRs. 
We can see, however, that attenuation of the lines (labelled with "RT") and background subtraction (labelled with "-bg") has a significant impact on the line ratios for both, the resolved (brown and grey lines) and the unresolved (blue and violet lines) cases.

We start with the [O III]/H$\mathrm{_\alpha}$ line ratio (row 1 of Fig.~\ref{fig:line_ratios}). 
The [O III] emission is sensitive to $T_e$, while H$\mathrm{_\alpha}$ depends more on $n_e$. 
As the primary shock (or reverse shock) travels into the neutral ambient medium (into the bubble centre), $T_e$ increases. Once the SNR starts to cool, the [O III] emission becomes bright in regions with relatively high $T_e$, i.e. behind the shock and inside the SNR bubble.
At the same time, the bubble expands and $n_e$ decreases, such that the H$\mathrm{_\alpha}$ emission increases more slowly than [O III]. It causes a peak in the [O III]/H$\mathrm{_\alpha}$ ratio along all LOS at around t$_{\rm evol}$ = 0.05 Myr and at t$_{\rm evol}$ = 0.2 Myr correspondingly. In Fig.~\ref{fig:pdf_sn_times}, we show that t$_{\rm evol}$ = 0.05 Myr corresponds to the transition time $t_{\rm TR}$ for gas with $\rho=10^{-23.5}\,{\rm g\, cm}^{-3}$, which in turn corresponds to the typical density of the MC material. 

From the [O III]/H$\mathrm{_\beta}$ line ratio (row 2 of Fig.~\ref{fig:line_ratios}), we can get the information about "complete" and "incomplete" recombination zones, respectively shell structure, for the SNR as explained in the theoretical works by \citet{Cox&Raymond1985, Hartigan1987} and observational works by \citet{Raymond1988, Boumis2005}.
Assuming a limit of around [O III]/H$\mathrm{_\beta}$ $\lesssim$ 6 (as derived in the aforementioned references), we can see that the shocks are completely recombined at the very beginning and at the end of the evolution.
Due to the complex density structure of the surrounding ISM, we can get high values of [O III]/H$\mathrm{_\beta}$ before \mbox{t$_{\rm evol}$ = 0.1 Myr} indicating an "incomplete" recombination zone. It is not so rare that high [O III]/H$\mathrm{_\beta}$ values are obtained in real observations, e.g. CTB 1 \citep{Fesen1997} or G 17.4-2.3 \citep{Boumis2002}.
We show that background subtraction (grey, brown, violet) significantly promotes the appearance of a "complete" recombination zone. Moreover, comparing the cases of a resolved SN with (grey) and without (brown) attenuation shows that attenuation is particularly important for H$\mathrm{_\beta}$ (also compare with Fig.~\ref{fig:flux_att}) and hence the [O III]/H$\mathrm{_\beta}$ ratio. Without attenuation, it has a value smaller than 6 during the whole time.

[N II]/H$\mathrm{_\alpha}$ (row 3) as well as [S II]/H$\mathrm{_\alpha}$ (row 4) are line ratios that help to classify an object as an SNR. 
Typical observed values for [N II]/H$\mathrm{_\alpha}$ lie in the range 0.5 - 1.0 \citep{Boumis2022}. 
We can classify our object as an SNR for resolved cases (brown and grey) for most of the time where optical emission is presented: from shortly after the explosion to the peak around t$_{\rm evol}$ = 0.05 Myr. The behaviour is comparable to the evolution of [O III]/H$\mathrm{_\beta}$ (see the description above). 
For resolved SNRs, attenuation does not play an important role in this case (grey and brown lines have almost the same values). On the other hand, for unresolved SNRs (blue and violet) the criteria to classify an object as an SNR is only fulfilled for one position of the observer (top view, violet-shaded region) after t$_{\rm evol}$ = 0.05 Myr. It means that for unresolved SNRs, it is rather difficult to trace the faint nitrogen line emitted when dense gas is shocked. 
This is consistent with high [N II]/H$\mathrm{_\alpha}$ line ratios being typically observed in older SNRs. 

[S II]/H$\mathrm{_\alpha}$ (row 4) is a typical tracer of shocked gas. 
Interestingly, for the cases of a resolved SNR without attenuation (brown) as well as a presumably unresolved SNR with background subtraction (violet), shock emission is identified for almost all times. 
However, both the unresolved case with attenuation (blue) and the resolved case with attenuation and background subtraction (grey) are mostly below the SNR limit.
It shows that attenuation can reduce the sulphur-bright thin shell emission. Hence, it could become impossible to properly identify an SNR in a denser ISM environment. 
At the same time, an accurate subtraction of the background for unresolved SNRs helps to boost the sulphur emission.

Overall, one can see from Fig.~\ref{fig:line_ratios} that weak lines (nitrogen and sulphur) 'benefit' from background subtraction in the sense that the line ratios are then shifted into the SNR zone.
This is not so critical for strong lines (oxygen and hydrogen).
However, it is worth noting that the whole analysis strongly depends on how well we can determine the abundance of different elements (i.e. the metallicity) in different parts of the SNR.

For all line ratios, at the end of the evolution of the SNR, the optical emission does not gradually fade yet, as some parts of the SNR are still "young" and more time is required for the whole remnant to merge with the surrounding ISM.

\subsection{BPT diagrams}\label{sec:BPT}

\begin{figure*}
	\includegraphics[trim={0.2cm 0.3cm 0.0cm 0.2cm},clip, width=15cm]{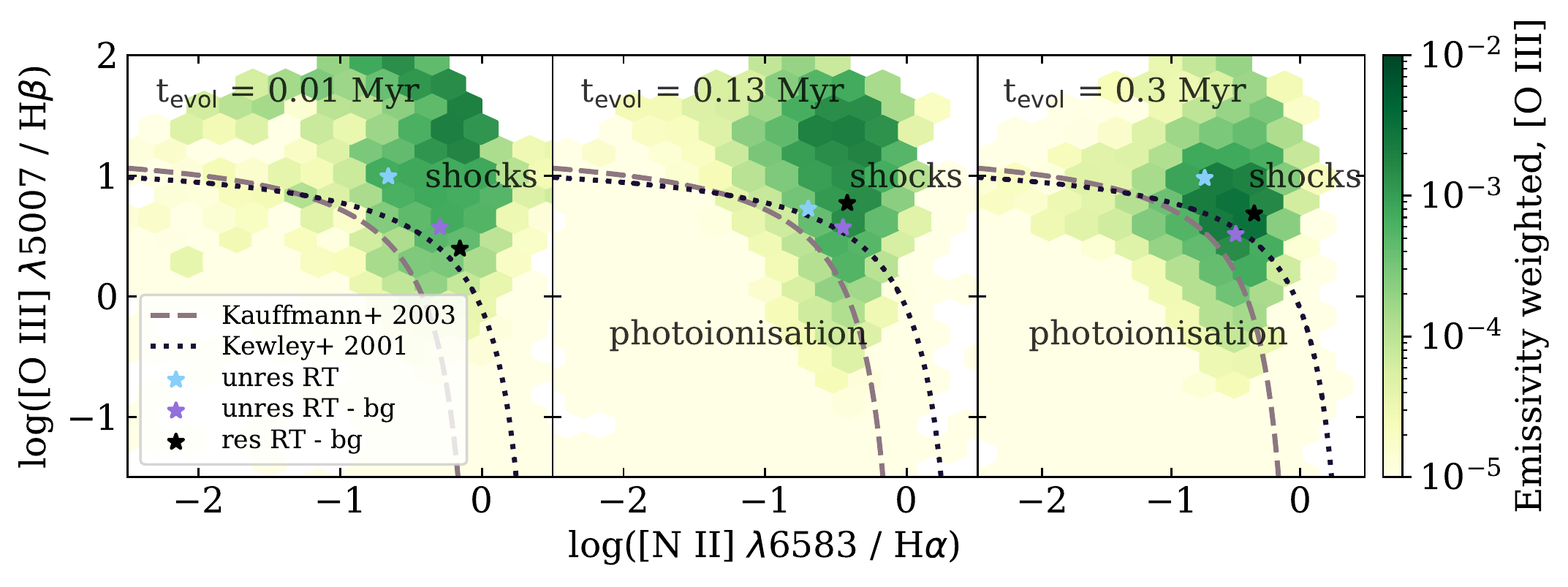}
    \label{fig:bpt_on}
    \vfill
    \includegraphics[width=15cm]{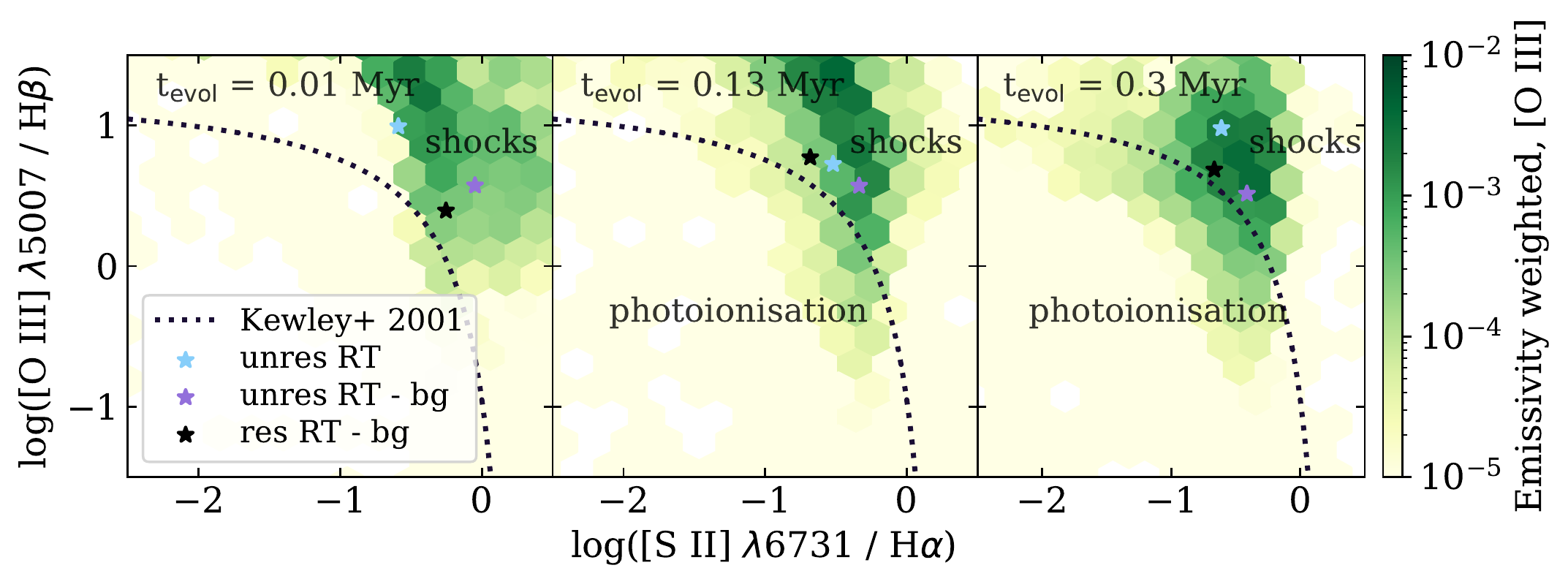}
    \caption{The classical BPT diagram (top row) and the sulphur BPT diagram (bottom row) for three different times (from left to right). The colour bar shows the line ratio distribution for the resolved SNR (with background subtraction and attenuation), weighted by the [OIII] emissivity. 
    The mean values for every calculation are star symbols; see the legend for the details. The reference lines classify the line ratios according to the main ionisation mechanism, i.e. photo-ionisation (lower left) or shock-ionisation (upper right). The dashed line is from \protect\citet{Kauffmann2003} and the dotted line is from \protect\citet{Kewley2001}.
    We find that the full distribution is clearly peaking in the shock-dominated regime, but the means are rather located in a "mixed region" (between the two regimes), in particular when the background has been subtracted.}
    \label{fig:bpt_os}
\end{figure*}

\begin{figure*}
	\includegraphics[trim={0.3cm 0.2cm 0.0cm 0.2cm},clip, width=15cm]{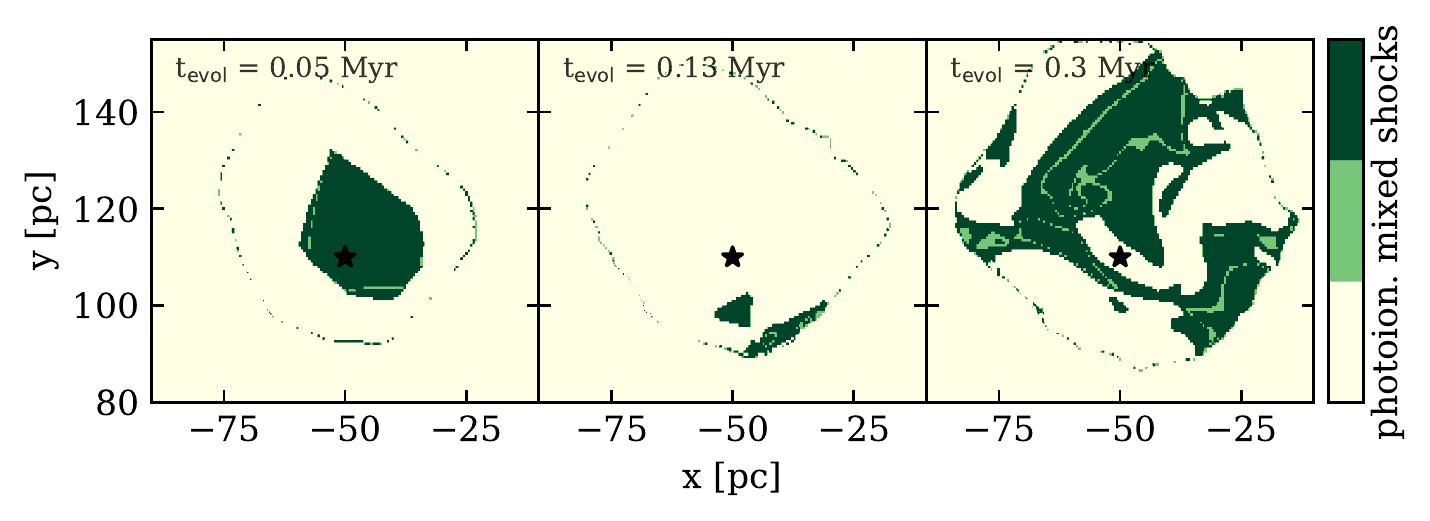}
    \caption{2D projected maps (along the $z$-axis) of the SNR at three different times (from left to right) colour-coded by the region where the local line ratio would lie within the classical BPT diagram shown in the top row of Fig.~\ref{fig:bpt_on}. The explosion centre is marked with a black star. We can see the outer shock and the inner bubble or the bubble surface, respectively. The line ratio map is very complicated due to the turbulent ISM with which the SNR interacts, which leads to a non-uniform cooling emission.}
    \label{fig:bpt_regions}
\end{figure*}

To investigate the distribution of the shocked gas by the SNR we use the BPT diagram. The BPT diagram is a way to classify the primary excitation source (photo-ionisation or shocks) of an object using observed line ratios. It is often used for unresolved objects such as entire galaxies. The classification mainly depends on the gas metallicity, the EUV field and some other parameters.
The BPT diagram compares the ratios of strong optical emission lines: [OIII]($\lambda 5007$)/$\mathrm{H \beta}$ and [NII]($\lambda6583$)/$\mathrm{H \alpha}$.

Fig.~\ref{fig:bpt_on} (top row) shows the BPT diagram computed for our SNR for three different times (from left to right: $t_{\rm evol}=0.01, 0.13, 0.3$~Myr). The colour-coded line ratio distributions are weighted by the [OIII]-emissivity and show the result derived for the resolved intensity maps (resolved case with attenuation and background subtraction), considering only pixels that have line intensities above the assumed observational threshold (Section~\ref{sec:tau}). For orientation, we also show two curves separating galaxies that are dominated by shocked emission (typically driven by AGN) and those dominated by H\textsc{ii} regions (photoionisation). The first line is based on purely theoretical work by \citet{Kewley2001} (using e.g., photoionisation, shock models, and population synthesis). The second line is derived from the analysis of data from the Sloan Digital Sky Survey \citep[SDSS;][]{SDSS, Stoughton2002} by \citet{Kauffmann2003}.

Note again that, in this work, we only account for collisionally excited emission. We do not include photoionisation driven by H II regions because the molecular cloud next to the SNR is not yet forming stars. Therefore, we can investigate the line emission contributed by the SNR without any contamination from nearby HII regions. We find that the HII-dominated part of the BPT diagram can be partially populated by emission from the SNR. It is likely that in a fully sampled patch of the galaxy with a multi-phase ISM, star formation and feedback \citep[such as, e.g. in][]{Rathjen2022} the lower left part of the BPT diagram would be dominated by real HII regions (see also Rathjen et al., in prep.). 

With time, the SNR line ratio distribution moves from the shock-dominated region to the so-called "mixed" or "composite" area: close to the boundary between the nominally H\textsc{ii} region-dominated and the shock-dominated emission. On the BPT diagrams that typically have a V-shape (due to the two main mechanisms of the excitation source), we are mostly reproducing the right part of the V, or the "shock wing" respectively \citep[][green-coloured background in Fig.~\ref{fig:bpt_on}]{Farage2010, Rich2014}, while we do not have a distinguishable "HII region" wing (left part of the V), which is expected since we do not include photo-ionisation, star formation, or stellar feedback in this study. This is consistent with previous works \citep{Allen2008, Kewley2019}.
We expect SNe to contaminate the classification of galaxies as they are populating both shock-dominated and mixed areas. Also, the line ratio distribution of an SNR depends on the SN evolutionary stage, i.e. on the age of the SNR as well as on the ambient gas distribution.

Additionally, we overplot several mean line ratios on the BPT diagrams (star symbols). For the resolved SNR with attenuation and background subtraction (black), the mean is the arithmetic mean of the pixel-based ratio of the corresponding individual emission line maps. We also show the means for the unresolved SNR with (violet) and without (blue) background subtraction and find that removing the background dramatically changes the resulting value: the mean is shifted downward (lower [OIII]/H$_{\beta}$) and to the right (higher [NII]/H$_{\alpha}$) so that it gets closer to the dividing line between the shock-dominated and the photo-ionisation dominated regimes. Mostly, the reason for this behaviour is that the emission of [O III] and both hydrogen lines are volume-filling (see Section~\ref{sec:shocks}) and thus, we typically have a lot of it in the background.
In case there are further ionizing sources in unresolved maps which are boosting the hydrogen lines, the mean values of the optical line ratios would further move towards the HII region part of the BPT, and hence SNRs could be missed entirely.

We also make use of the [SII] $\lambda 6731$/$\mathrm{H \alpha}$ ratio to plot the sulphur BPT diagram in Fig.~\ref{fig:bpt_os} (bottom row).
In a photo-ionised region, some UV photons have sufficiently high energy ($\sim$ 23.3 eV) to doubly ionise sulphur (S$^{++}$). In addition to that, photo-ionised regions also contain some amount of collisionally excited S$^{+}$. In SNRs the situation is the opposite: due to the large recombination region behind the shock, the SNR will mainly contain S$^{+}$ rather than S$^{++}$.
This is why the [SII]($\lambda6731$)/$\mathrm{{H \alpha}}$ line ratio is a standard diagnostic tool to detect SNRs. The sulphur BPT diagram also contains the \citet{Kewley2001} separation line.
During the whole time evolution of the SNR, our mean values are clearly located in the shock-dominated region. This confirms the possibility of correctly classifying an SNR on the sulphur BPT, even for unresolved objects.

Recently, many attempts have been made to improve the BPT diagrams using other optical line ratios (creating a multi-dimensional optical line ratio space) and modern machine learning techniques to better classify observed objects or to better constrain their physical conditions \citep[e.g.][]{Vogt2014, Kewley2019, Stampoulis2019, Ho2019, Zhang2020, Ji2020, Rhea2021}.
However, none of the new models is a universal tool that could be used in both theoretical work and observations.
Therefore, despite all these advances in methods, we need more physical models that use BPT diagrams to understand what is driving the excitation of the emitting gas. We will investigate this aspect in future studies using machine learning with a larger dataset of SNRs \citep[Smirnova et al, in prep., based on the full dataset of][]{Seifried18}.

To visualise the location of pixels with line ratios located in either the shock-dominated (dark green), the mixed (green), or the photo-ionised (yellow) regimes of the BPT diagram, we plot the 2D projected map (projection along the $z$-axis) in Fig.~\ref{fig:bpt_regions} for the same time steps as above (from left to right). It is interesting to note that the mixed region follows the outer shock position (which is shown in Section~\ref{sec:shocks}), which is consistent with theoretical predictions for the position of SNRs in the BPT diagram \citep{Allen2008, Kewley2019}. The shock-dominated regions mainly follow the hotter gas in the bubble centre and strongly cooling regions on the bubble surface that emit in the EUV and X-ray. All other parts of the SNR bubble are indistinguishable and appear in the photo-ionised regime although they are still shock-ionised in our case.

\subsection{Physical conditions in the SNR derived from simulations and emission line ratio diagnostics} \label{sec:sulphurratio}

\begin{table}
    \centering
    \begin{tabular}{c|c|c|c|c|c|c}
        \hline
        t$_{\rm evol}$ & \multicolumn{4}{c|}{SN box} & \multicolumn{2}{|c|}{Best fit model} \\
        \hline
        & total & total & [S II] & [O III]  &&\\
        & n$_{\rm e}$ & T$_{\rm gas}$ & n$_{\rm e}$ & T$_{\rm gas}$ & n$_{\rm e}$ & T$_{\rm e}$ \\
        Myr & [cm$^{-3}$] & [K] & [cm$^{-3}$] & [K] & [cm$^{-3}$] & [K] \\
        \hline
        0.01 & 40 & 949 & 981 & 66 621 & 1100 & 85 500 \\
        0.13 & 42 & 2234 & 390 & 88 006 & 500 & 95 000\\
        0.3 & 45 & 12386 & 110 & 99 871 & 270 & 108 000 \\
        \hline
    \end{tabular}
    \caption{Comparison of electron density (n$_{\rm e}$) and electron temperature (T$_{\rm e}$) for different t$_{\rm evol}$ (first column). 2d and 3d column computed for the SN box (corresponds to Tab.~\ref{tab:boxes}, columns 2,3,4); 4th column computed for [S II] emission region only in SN box;  5th column computed for [O III] emission region only in SN box and 6th, 7th columns result from the {\sc PyNeb}}
    \label{tab:neTe}
\end{table}

Optical line ratios are used to determine the physical parameters of observed SNRs.
First, one may estimate the electron temperature ($T_{\rm e}$ ) and the electron density ($n_{\rm e}$ ) from the [O III] ($\lambda 4363$)/[O III] ($\lambda 5007$) and [SII] ($\lambda 6731)/ (\lambda 6717$) line ratios. The sulphur line ratio is almost independent of $T_{\rm e}$ \citet{Osterbrock2006} and the oxygen line ratio depends only on $T_{\rm e}$. That is why we can define $T_{\rm e}$ and $n_{\rm e}$ reliably. The sulphur line ratios from our post-processing are presented in Table~\ref{tab:line_ratios_res} and fall in the range from 0.2 to 1.0. A typical value for  old (evolved) SNRs in the low-density medium is about 1.1-1.5, \citep{Frank2002}). It is worth noting, that our SNR is in a denser medium where the values of [S II] ($\lambda 6731$)/[S II]($\lambda 6717$) around 1.0 are common to observe.

We use the {\sc PyNeb} tool \citep{Luridiana13} - a \textsc{Python} package which computes emission lines using an $n-$level atom approach. It solves the equilibrium equations for an n-level atom to determine the level populations (from which the emissivities can be calculated). 

Using the above-mentioned oxygen and sulphur line ratios, we obtain $n_{\rm e} \approx 200-1000 $ cm$^{-3}$ and  $T_{\rm e} \approx 85 500 - 108 000 $ K. The electron density indicates a cooling post-shock recombination layer. The electron temperature corresponds to a hot interior of the SNR bubble (as we mentioned before, see Fig.~ \ref{fig:2phase_diagram}). 
The time evolution looks like this: we start with a high electron density (about 1100 cm$^{-3}$) and $T_{\rm e} \approx 85 500$ K), which is typical for young SNRs, and end up at a number density $n_{\rm e} \approx$ 270 cm$^{-3}$ and an electron temperature $T_{\rm e} \approx 10.8 \times 10^4$ K.

To verify the line ratio-based results, we determine our electron number densities $\mathrm{n_e}$ from the simulations for each cell using:
\begin{equation}
   n_\mathrm{e} = \frac{\sum I_{\mathrm{cell}} \times \rho_{\mathrm{cell}} \times x_{\mathrm{H^+}} / m_\mathrm{p}} {\sum I_{\mathrm{cell}}},
\end{equation}
where $x_{\rm H^+}$ is the fractional abundance of ionised hydrogen, $I_{\rm cell}$ is the S$^+$ intensity in the cell, $m_{\rm p}$ is the mass of a proton.
The result values for  $n_{\rm e}$ are summarised in Table~\ref{tab:neTe}.
The $n_{\rm e}$ found in the simulation (from the area where we have sulphur emission in simulations, fourth column) is similar to the one derived using the line ratios and {\sc PyNeb} (sixth column).
Unfortunately, we cannot calculate $T_{\rm e}$ directly from the simulations. 
But $T_{\rm e}$ (from the line ratios, seventh column) and $T_{\rm gas}$ (from the area where we have oxygen emission in simulations, fifth column) have no great differences for the part of the SNR bubble that we compare because we assume thermal equilibrium between the electron and gas temperature. In a conclusion, this method works quite well in describing the mean electron densities (error: 11-40 per cent) and temperatures (error only: 8-20 per cent) despite the variations along the line of sight and between lines of sight. Our best-fit model seems to always be biased toward overestimating $n_{\rm e}$, $T_{\rm e}$ and this overestimation seems to grow with time (ending with the biggest error of 40 and 20 per cent correspondingly).

Combining our results for the electron density with the fact that the SNR has a diameter of approximately 25 -- 55~pc (the minimum and maximum diameter of the SNR bubble during the $t_{\rm evol}$, see Fig.~\ref{fig:t_d_bands}), one may calculate the energy of the explosion using the following equation from \citep{McKee75}: 
\begin{equation}
    E = 5.5 \times 10^{43} n_\mathrm{e} D^3~\mathrm{erg},
\end{equation}
where $D$ is the SNR diameter in parSection
This results in $E \approx 8.6 \times 10^{50} - 1.8 \times 10^{51}$ erg.
These values are comparable to the actual SN explosion energy in our simulation of $E_{\rm SN}=10^{51}$~erg.
Still, the difficulty of measuring a well-defined diameter is a substantial source of inaccuracy in this calculation (after all, our SNR is not spherically symmetric, as can be seen in Fig.~\ref{fig:t_d_bands}).

In general, we have quite a good agreement between simulations and optical line ratios numbers.
However, $T_{\rm e}$ and $n_{\rm e}$ are changing notably during the simulation time (especially in different parts of the bubble that do not evolve simultaneously), so more statistical data are needed to draw a conclusion on the accuracy of the method.

\subsection{Detection and analysis of shocks}
\label{sec:shocks}
Shock waves are essential components of young SNRs: for most, the shell structure is defined by the forward shock which moves into the ISM and the reverse shock which moves back towards the centre and into the ejecta.

\begin{figure*}
    \includegraphics[width=15cm]{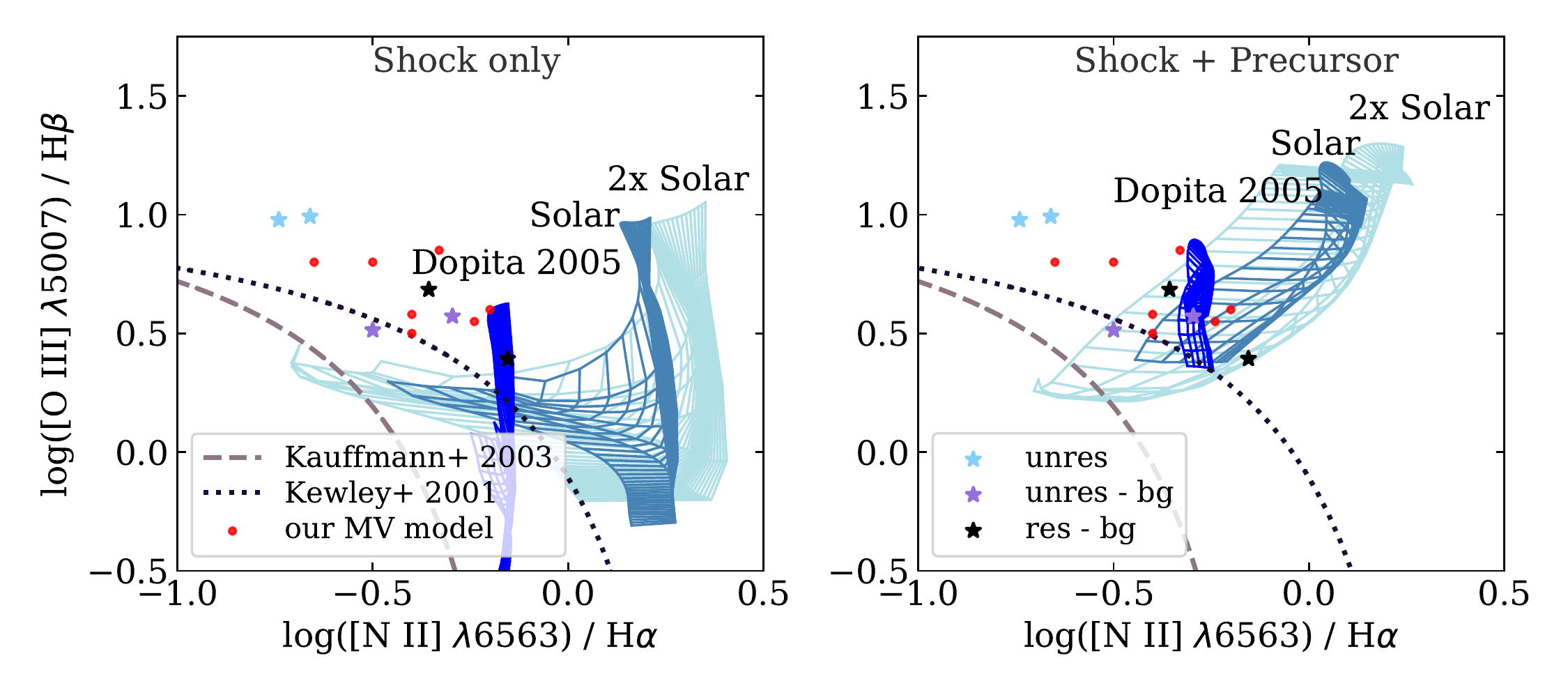}
    \caption{Shock diagnostic diagram based on the same optical line ratios as BPT diagram. Different abundances (showed in blue colours) cover shock velocities ranging from $200$ to $1000$ $\mathrm{km}$ $\mathrm{s^{-1}}$ (an increase from top to the bottom for each grid, one line - one velocity;), preshock density of n$_{\rm 0}$ = 1  $\rm cm^{-3}$, magnetic parameter between $10^{-4}$ to 10 $\mu$G $\mathrm{cm^{3/2}}$ (increase for each abundance set from left to right). Note that the plot displays the line ratios from the shocked gas only.
    The star symbol and the arrow with the corresponding colour show the mean at t$_{\rm evol}$ = 0.01 Myr and t$_{\rm evol}$ = 0.3 Myr. The colours are the same as in Fig.~\ref{fig:bpt_on}. Red stars are shock models calculated with {\sc MAPPINGS V} with our abundances (see Tab.~\ref{tab:metalls}) with different initial parameters, similar to simulations. The shock+Precursor model works better for our optical line ratios, however, different abundances (compared to the default ones, shown in the figure) are required to fit the obtained data.}
    \label{fig:shock}
\end{figure*}

We use 1D radiative shock models to estimate the shock velocity and their magnetic field strength.
In Fig.~\ref{fig:shock}, we compare our mean line ratios (at $t_{\rm evol}$ = 0.01 Myr and $t_{\rm evol}$ = 0.3 Myr, and we show three different cases in black, blue, and violet) with a grid of models from \citet{Alarie2019}. Note that the shock models are interpreted on the basis of the same line ratios as in the BPT diagrams (see Fig.~\ref{fig:bpt_on} and Section~\ref{sec:BPT}). Additionally, we show the same reference lines (grey dashed and dotted lines) as in Fig.~\ref{fig:bpt_on}. We find that, for the unresolved case (blue stars), the background subtraction (leading to the violet stars) substantially changes the shock parameters which would be determined from the comparison with the shock models.

All our mean line ratios are quite offset from the grid of shock models, so it is essential to mention that the shock diagram is extremely sensitive to the initial abundance of the environment (mainly nitrogen).
Of course, one cannot quantitatively define the time-dependent metal enrichment of the ISM from the SN itself from a real observation, so one should be careful when interpreting the results based on the shock diagram.
The red dots show additional shock models that we calculate with {\sc MAPPINGS V} for our specific metal abundances as given in Tab.~\ref{tab:metalls} and for different gas densities, temperatures, and magnetic field strengths that are better suited to our physical conditions than the given model grids. In particular, we consider a temperature range from 10$^{7}$ to 10$^{4}$ K (see Fig.~\ref{fig:2phase_diagram}) and magnetic field strength from 10$^{-4}$ to 10 $\mu$G.
Mostly, the gas metallicity (and temperature) is responsible for the different line ratios of the newly computed models.

In {\sc MAPPINGS V}, it is possible to either use the shock model (left panel) or the shock+precursor model (right panel).
Typically, the post-shock regions (of radiative or non-radiative shocks) are sources of ionising photons.
When these regions cool and recombine a continuum flux is produced.
It propagates further upstream and forms a photoionisation precursor \citep{Cox1972, Raymond1991}.
Because our precursor gas is pre-ionised, the model with the shock+precursor fits better. 

\begin{figure*}
	\includegraphics[width=13cm]{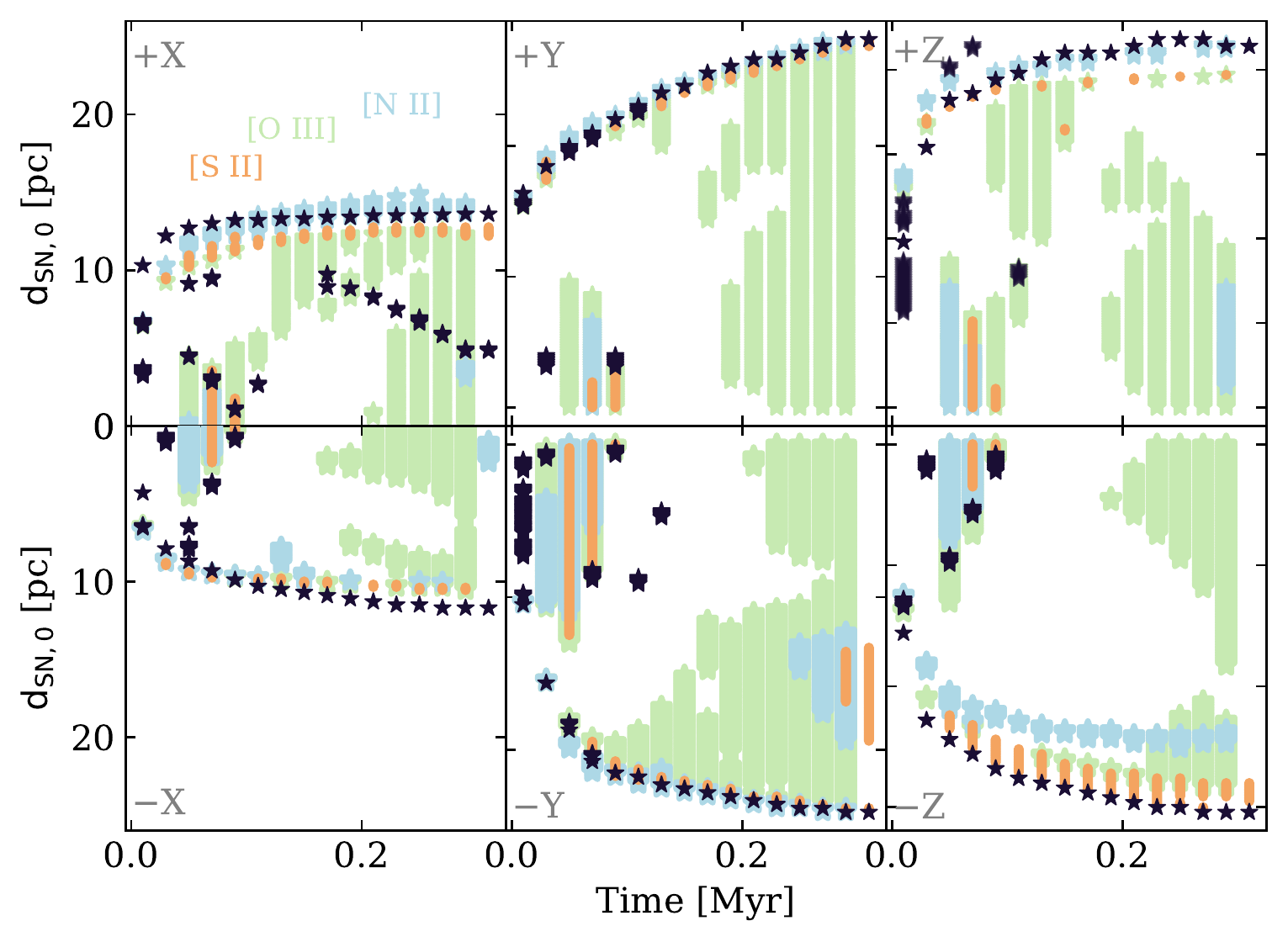}
    \caption{Time evolution of the radiation position relative to the centre of the SN explosion along each axis.
    [S II] emission (orange), [O III] emission (light green), [N II] emission (light blue) and shock position (black stars).
    Not all shock wave positions were calculated using the shock finder.
    The missing points were calculated based on the temperature jump.
    The shock velocity also can be calculated from these data easily.
    We start from 800-1000 km/s during the first time steps (0.01 Myr) and then the velocity decreases dramatically to 200-400 km/s (around 0.05 Myr).
    At the end of the time evolution (0.3 Myr) the shock velocity is almost constant around 12-50 km/s.
    Note that the scatter in velocity is big because it depends on the environmental density.
    In some directions, shock expands freely (+Y or -Y) and in another, it hits the MC and almost stops (+X or -X).}
    \label{fig:shock_pos}
\end{figure*}

Further, we find that the best fit is with models that have shock velocities of $350-500$ km s$^{-1}$ and a weak magnetic field (for the right panel) with $\rm{Z = 1-2\ Z_\odot}$ (where $ Z_\odot$ is solar metallicity).
This is a reasonable estimate of the actual shock velocity ($200-400$ km s$^{-1}$) at time $\mathrm{t_{evol}} = 0.05$ Myr or for a mean shock velocity (see also Fig.~\ref{fig:shock_pos} for more details). The shock velocity decreases over time in the simulation from 2400~km~s$^{-1}$ to 25~km~s$^{-1}$. 
The weak magnetic field strength is also consistent with our simulations, as we initially set B = 3~$\mu$G in the simulation (see Section~\ref{sec:sim}).

We implemented a shock-detection routine \citep[] [based on]{Lehmann2016} that can identify and extract shock regions in the SNR, as shown in Fig.~\ref{fig:shock_pos}.
 As theoretically predicted, the emission of optical lines does not exactly track the positions of the shock front.
 The optical emission is generated mainly in the cooling post-shock region, which is visible in Fig.~ \ref{fig:shock_pos}.
 However, we can trace the approximate shock position using [S II] (in orange, as sulphur is collisionally exited). It can even capture the position of the reverse shock. Nitrogen ([N II]) is not so sensitive to the shock position: the emission can be spatially separated from the shock position due to the lower excitation temperature. [O III] can trace the main shape of the SNR bubble but does not correlate with the shock position anyhow.
 It is clearly seen which areas of the supernova bubble are occupied by different optical lines: [O III]] fills the bubble and appears where the gas has already cooled down enough, while the [S II] and [N II] radiation is a thin layer after/before the shock wave.
 Thus, even if we do not detect the position of the shock wave perfectly using [S II] as a tracer, we can still use it as a good approximation.
In addition, from an estimate of the thickness of the [S II] layer, it can be concluded that a simulation resolution of less than one parsec is required to resolve the radiation of cooling SNR shock waves in the optical band.
 The evolution of the main and reverse shock waves as a 3D projection and on a 2D slice from our simulations are shown in Appendix~\ref{appendix:shock}.

\section{Conclusions}
\label{sec:conclusions}

We develop a post-processing module for the {\sc FLASH} code, which uses the updated collision data from {\sc MAPPINGS V} to produce realistic emission maps of simulated SNRs.

We show that a significant amount of cooling energy is released through EUV, FUV, and Optical radiation.
Because of the complex and turbulent structure of the surrounding ISM, the characteristic evolution times (from the duration of the Sedov-Taylor phase to the transition and PDS phases) of the SNR can differ greatly for different parts of the SNR bubble. The SNR bubble itself is highly asymmetric, leading to significant projection effects. 

We produce optical emission line maps ($\mathrm{H{\alpha}}$, $\mathrm{H{\beta}}$, [O III] ($\lambda 5007)$, [S II] ($\lambda 6717, 6731$), and [N II] ($\lambda 6583$) that show the same features as in real observations of SN remnants interacting with the surrounding ISM. 
We investigate the following optical line ratios: [O III] ($\lambda 5007)/ \mathrm{H{\alpha}}$, [S II] ($\lambda 6717$)/($\lambda 6731$),  [S II] ($\lambda 6731$)/$\mathrm{H{\alpha}}$, [N II] ($\lambda 6583$)/$\mathrm{H{\alpha}}$, [O III] ($\lambda 5007$)/$\mathrm{H{\beta}}$ that we also find to be mostly consistent with observations.
We investigate where our SNR appears in the classical and the sulphur BPT diagrams and show that it moves with time from the shock-dominated region to the mixed region (between the regimes of shock-dominated and classical H{\sc ii} regions, which are actually not included here). This behaviour is mainly caused by shock dissipation in the turbulent ISM. 
For unresolved SNRs removing the background completely changes the position on the BPT diagram: if there are further ionizing sources in unresolved maps (boosting the hydrogen lines), the SNR position on the BPT diagram would move towards the photoionisation part, and hence SNR could be missed entirely.

Furthermore, using sulphur lines, we calculate an electron density $n_{\rm e} \approx$ 200 - 1000 cm$^{-3}$ and an electron temperature $T_{\rm e} \approx 85 500 - 10.8 \times 10^{4}$~K from the oxygen emission.
These values are consistent with typical SNR observations and our simulation data. Yet, they indicate the presence of different parts of the SNR: a cooling post-shock recombination layer is traced by the sulphur emission, and the hot bubble is traced by the oxygen emission.
Using the calculated electron density, we estimate the initial supernova energy and find values of $E \approx 8.6 \times 10^{50} - 1.8 \times 10^{51}$ erg, which agrees fairly well with the injected explosion energy of $E_{\rm SN} = 10^{51}$ erg.
The variation in $E$ arises since the size of the SNR bubble cannot be determined exactly due to the asymmetry of the surrounding turbulent ISM and depends on the evolutionary stage.

In addition, we use shock models to estimate the shock velocity and find values between $350 - 500~\mathrm{km \;s^{-1}}$, which are in agreement with the mean velocity calculated from the synthetic observations. For the first time, we can very well reproduce the position of the optical emission and shock waves. We can clearly see the absence of spherical symmetry in both forward and backward shock waves.
It is shown that different ions are formed in other regions under different initial conditions.
Overall, in order to properly interpret the results, we can show that it is vital to consider the projection effect, a realistic density distribution of the surrounding ISM, and the most accurate metallicity estimate available.
All of these factors significantly affect the accuracy of optical line diagnostics.

Finally, accounting for cooling radiation self-consistently in simulations is extremely important throughout the entire evolution of the SNR. It can significantly change the entire dynamics of the remnant's evolution.

\section*{Acknowledgements}

The authors thank the anonymous referee for the fast reviewing process. We also would like to thank Kathryn Kreckel and her group for useful discussions and information about optical observational limits. E.I.M., S.W. and S.D.C. acknowledge funding by the European Research Council through ERC Starting Grant No. 679852 'RADFEEDBACK'. S.W., D.S., T.-E.R. acknowledge support by the German Science Foundation (DFG) via SFB 956, projects C5 and C6. T.-E.R. and D.S. acknowledge funding from the programme "Profilbildung 2020", an initiative of the Ministry of Culture and Science of the State of Northrhine Westphalia. The sole responsibility for the content of this publication lies with the authors. 
The software used in this work was partly developed by the DOE NNSA-ASC OASCR Flash Centre at the University of Chicago \citep{Fryxell00, Dubey08}. The following {\sc Python} packages were used: {\sc NumPy} \citep{Numpy}, {\sc SciPy} \citep{SciPy}, {\sc Matplotlib} \citep{matplotlib}, {\sc yt} \citep{yt}. We also used {\sc Paraview} for the 3D visualisation \citep{ParaView}.

\section*{Data Availability}

The data underlying this article will be shared on reasonable request to the corresponding author. The CESS module is available on GitHub.



\bibliographystyle{mnras}
\bibliography{ref} 




\appendix

\section{Volume weighted PDF for the SNR box}
\begin{figure*}
	\includegraphics[width=16cm]{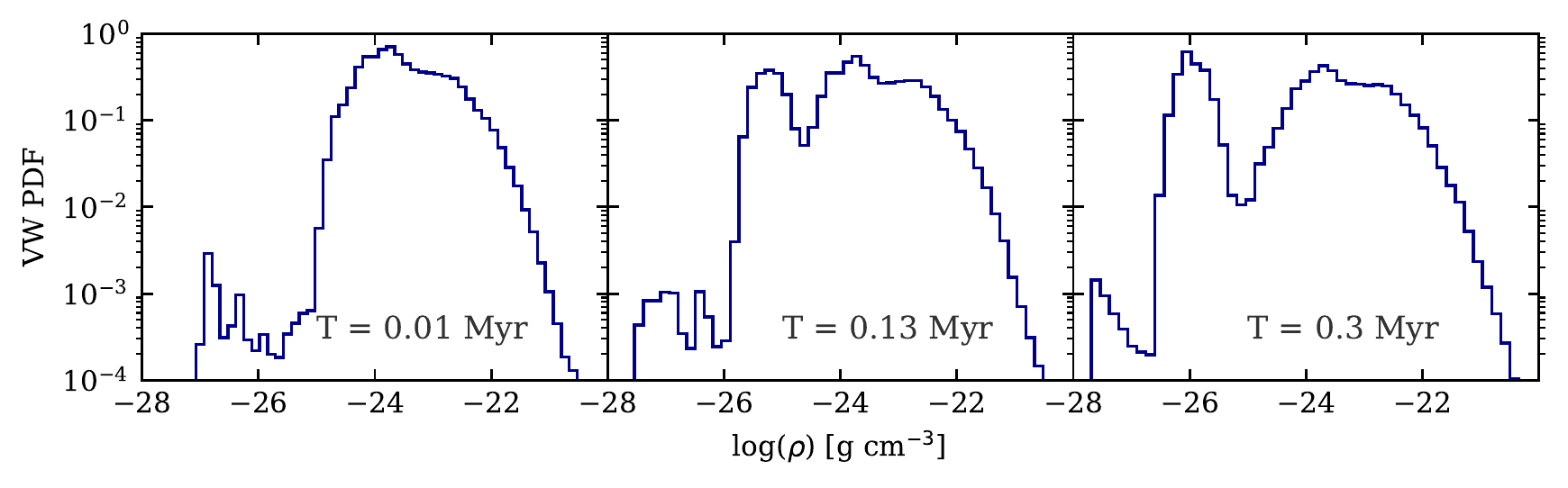}\\
    \caption{Time evolution (from left to right) of the volume-weighted PDF of the SNR box.}
    \label{fig:vw_pdf_evol}
\end{figure*}

From a volume-weighted PDF time evolution of SNR box only (left panel, Fig.~\ref{fig:energy_3Dsim}) on Fig.~\ref{fig:vw_pdf_evol} we can see how the distribution of density is changing while the SNR interacts with the MC. 
At the very beginning, we have most of the gas in a dense phase with the peak around 10$^{-23}$ g cm$^{-3}$, but after 0.13 Myr there is already a second peak at 10$^{-25}$ g cm$^{-3}$, due to the dispersion of the MC with shock, which moves to 10$^{-26}$ g cm$^{-3}$ at time 0.3 Myr. 
That is why, it is far from a realistic view to use any unique physical time scales for the whole SNR simulation box, which typically uses the approach of the uniform density ISM.

\section{Lines emission maps and line ratios}
\label{appendix:emission}
This section provides examples of 2D emission maps in different optical emission lines after the post-processing procedure for three consecutive time points. One can see the difference between the volume-filling radiation and the radiation at the edge. In addition, emission projection effects are visible: increased radiation from the supernova bubble regions in the middle panel of all pictures and an overlay of instabilities from different planes in the upper right side of the bubble. Note that here background emissions have not been subtracted for demonstration purposes. Previous SNRs are visible in almost every image in the upper right corner.

\begin{figure*}
\vspace{-5pt}
	\includegraphics[width=16cm]{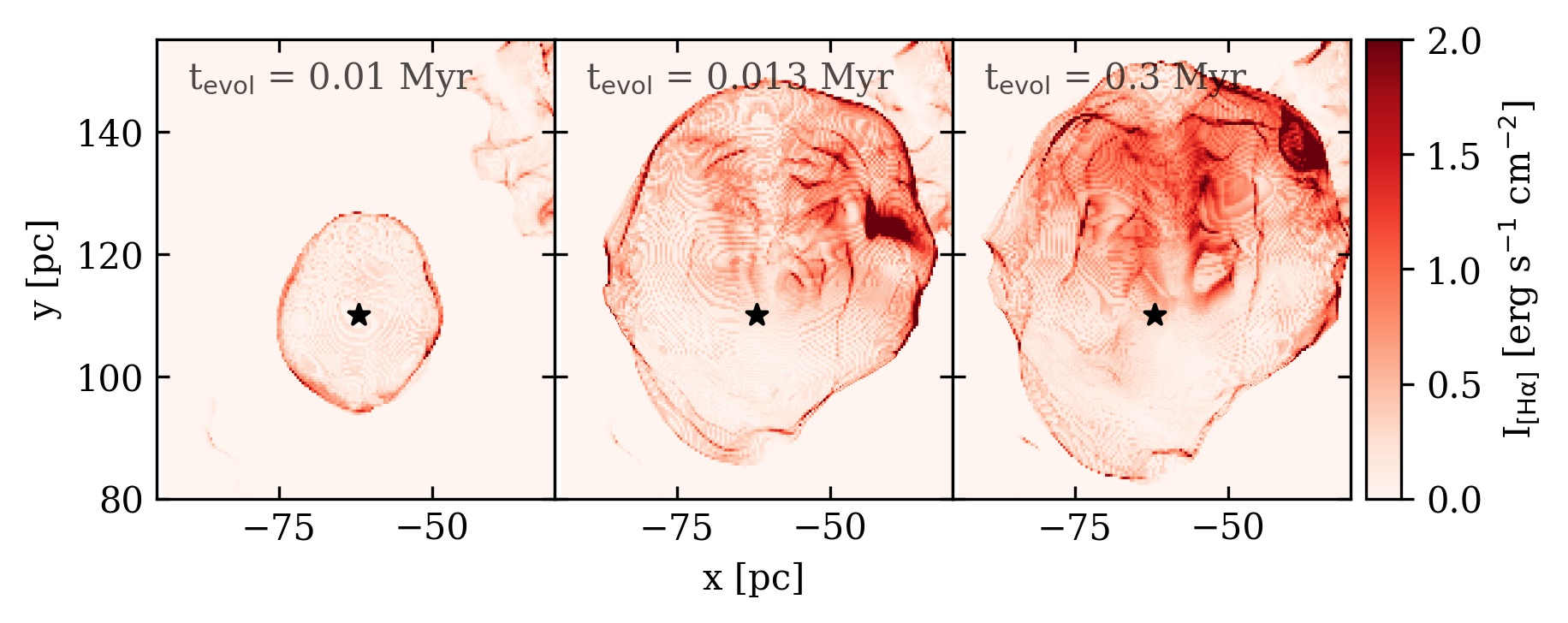}\\
    \label{fig:ha_row}
\vspace{-22pt}
\vfill
	\includegraphics[width=16cm]{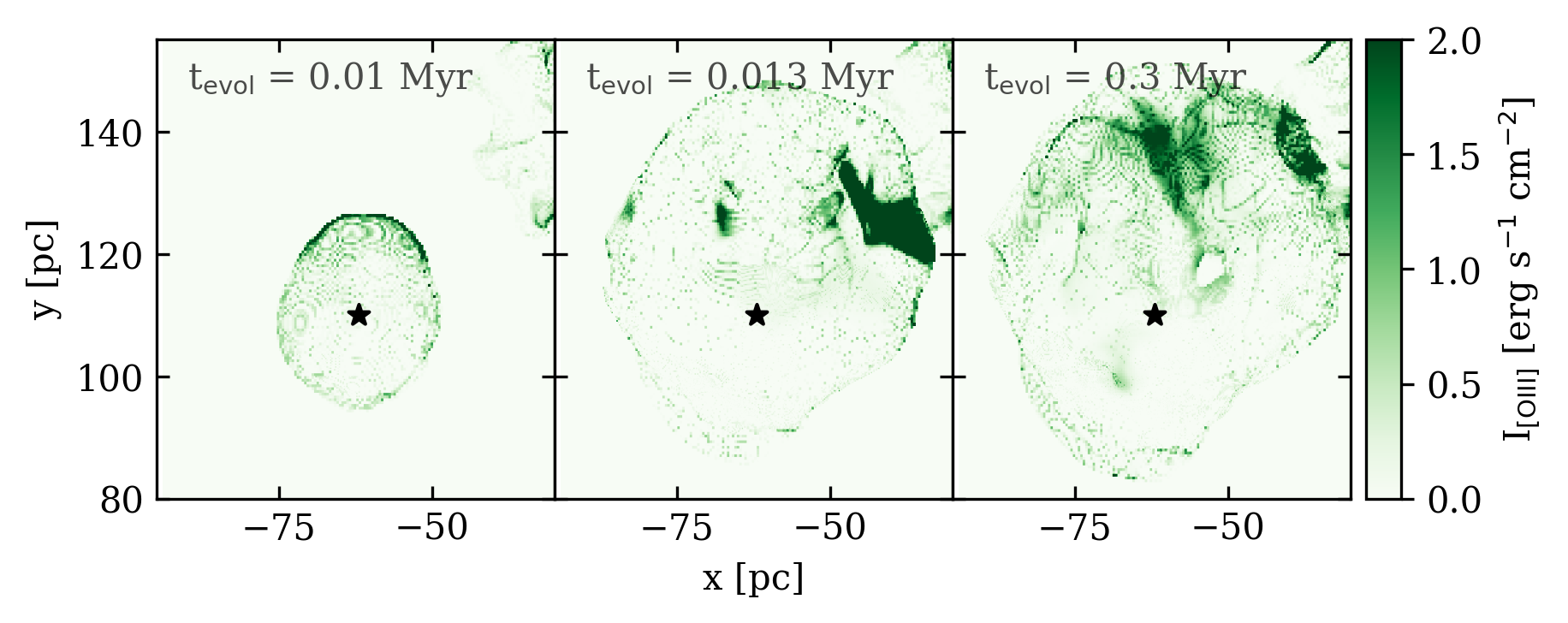}\\
    \label{fig:o3_row}
\vspace{-22pt}
\vfill
	\includegraphics[width=16cm]{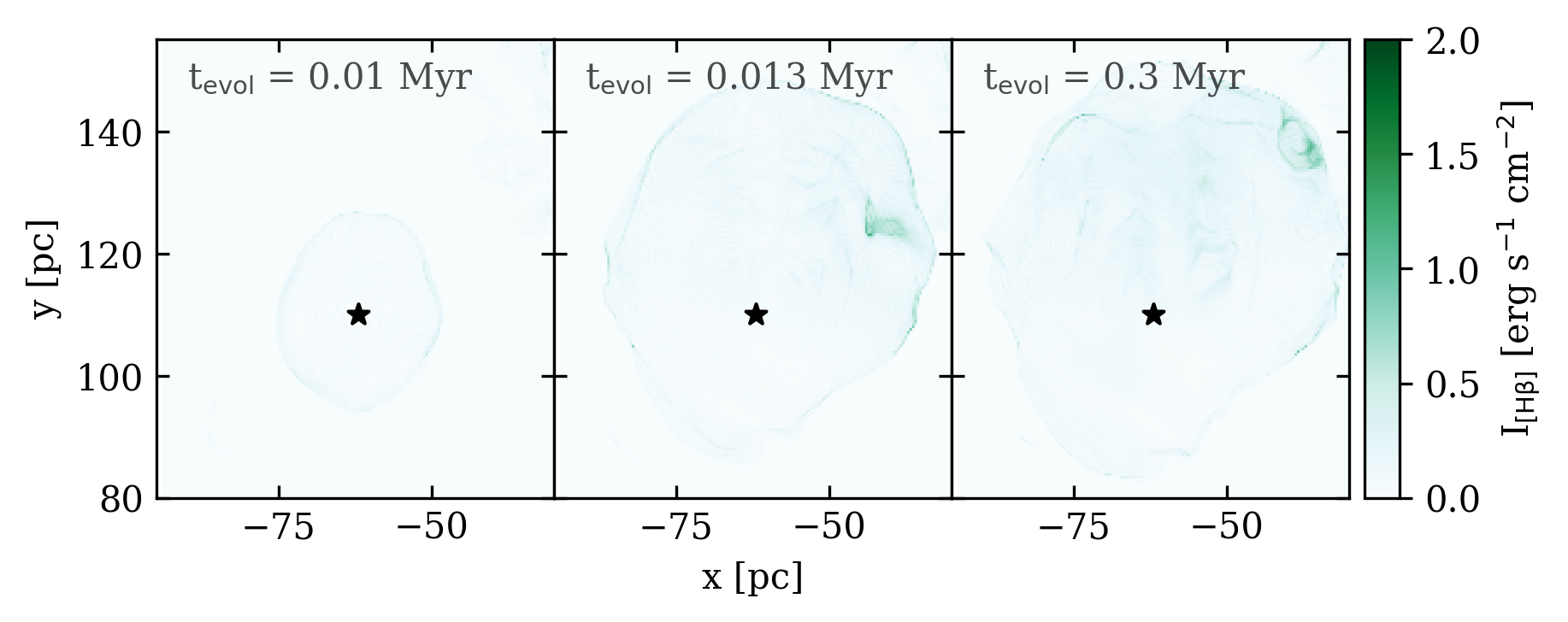}\\
    \label{fig:hb_row}
\vspace{-22pt}
\vfill
	\includegraphics[width=16cm]{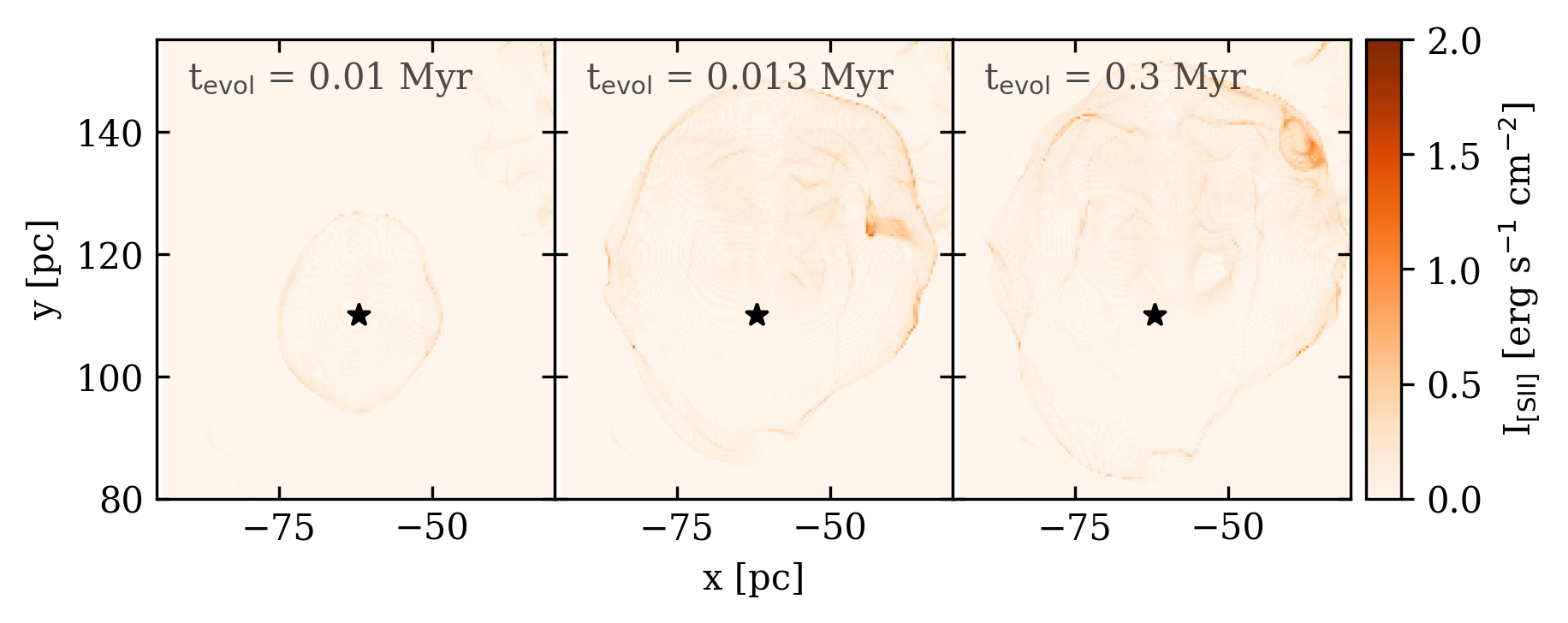}\\
    \caption{Time evolution (from left to right) of the [H $\mathrm{\alpha}$] $\mathrm{\lambda 6563}$, [O III] ($\mathrm{\lambda 5007}$), [H $\mathrm{\beta}$] $\mathrm{\lambda 4861}$, [S II] $\mathrm{\lambda 6731}$ emission (from top to bottom). Intensity is normalised to the value $10^{-17}$.}
    \label{fig:s2_row}
\end{figure*}

\begin{table*}
    \centering
    \begin{tabular}{ccccccll}
    \hline
    Obtained value & $[$O III] ($\lambda 5007$)/$\mathrm{H{\beta}}$ & $[$O III] ($\lambda 5007)/\mathrm{H{\alpha}}$ & $[$N II] ($\lambda 6583$)/$\mathrm{H{\alpha}}$ & $[$S II] ($\lambda 6731$)/$\mathrm{H{\alpha}}$ & $[$S II] ($\lambda 6717$)/($\lambda 6731$)\\
    \hline
    -bg x-axis & [0.6, 2.5] & [0.8, 6.1] & [0.4, 0.7] & [0.5, 0.6] & [0.6, 1.0] \\
    RT-bg x-axis & [1.9, 22.1] & [0.8, 7.5 ]& [0.4, 0.7] & [0.2, 0.3] & [0.4, 0.9]  \\
    -bg y-axis & [0.6, 2.8] & [0.7, 7.0] & [0.4, 0.7] & [0.3, 0.6] & [0.5, 0.7]\\
    RT-bg y-axis & [1.8, 23.1] & [0.6, 8.3] & [0.4, 0.7] & [0.2, 0.8] & [0.2, 0.7]\\
    -bg z-axis & [0.6, 2.5] & [0.6, 6.1]  & [0.4, 0.7] & [0.3, 0.6]  & [0.6, 1.0]\\
    RT-bg z-axis & [1.9, 22.9] & [0.6, 6.7] & [0.4, 0.7] & [0.2, 0.7] & [0.5, 1.0] \\
    \hline
    \end{tabular}
    \caption{Minimum and maximum value of line ratios derived from synthetic observations in the optical band for resolved SNRs. These values are calculated taking into account two possible positions of the observer (top or bottom view of the cube). We also marked in the first column, along which axis the projection was made.}
    \label{tab:line_ratios_res}
\end{table*}

\begin{table*}
    \centering
    \begin{tabular}{cccccll}
    \hline
    Obtained value & $[$O III] ($\lambda 5007$)/$\mathrm{H \beta}$ & $[$O III] ($\lambda 5007)/\mathrm{H \alpha}$ & $[$N II] ($\lambda 6583$)/$\mathrm{H \alpha}$ & $[$S II] ($\lambda 6731$)/$\mathrm{H \alpha}$ \\
    \hline
    RT x-axis & [2.0, 22.1] & [0.4, 6.7] & [0.1, 0.5] & [0.2, 0.3]  \\
    RT-bg x-axis & [5.3, 98.8] & [0.4, 5.8 ]& [0.2, 0.7] & [0.2, 0.6]  \\
    RT y-axis & [1.8, 23.1] & [0.4, 6.8] & [0.2, 0.5] & [0.2, 0.4]\\
    RT-bg y-axis & [9.3, 75.5] & [0.4, 5.8] & [0.1, 0.7] & [0.2, 0.9] \\
    RT z-axis & [1.9, 22.9] & [0.4, 7.1]  & [0.1, 0.5] & [0.2, 0.3] \\
    RT-bg z-axis & [4.8, 82.5] & [0.4, 6.9] & [0.3, 0.7] & [0.2, 0.6] \\
    \hline
    \end{tabular}
    \caption{Minimum and maximum value of line ratios derived from synthetic observations in the optical band for unresolved SNRs. Note that we do not have here a close sulphur line ratio, because these lines are not resolved for these types of objects. See other details in Table \ref{tab:line_ratios_res}.}
    \label{tab:line_ratios_unres}
\end{table*}

\section{Shock cells and density projection}
\label{appendix:shock}
\begin{figure*}
	\includegraphics[width=14cm]{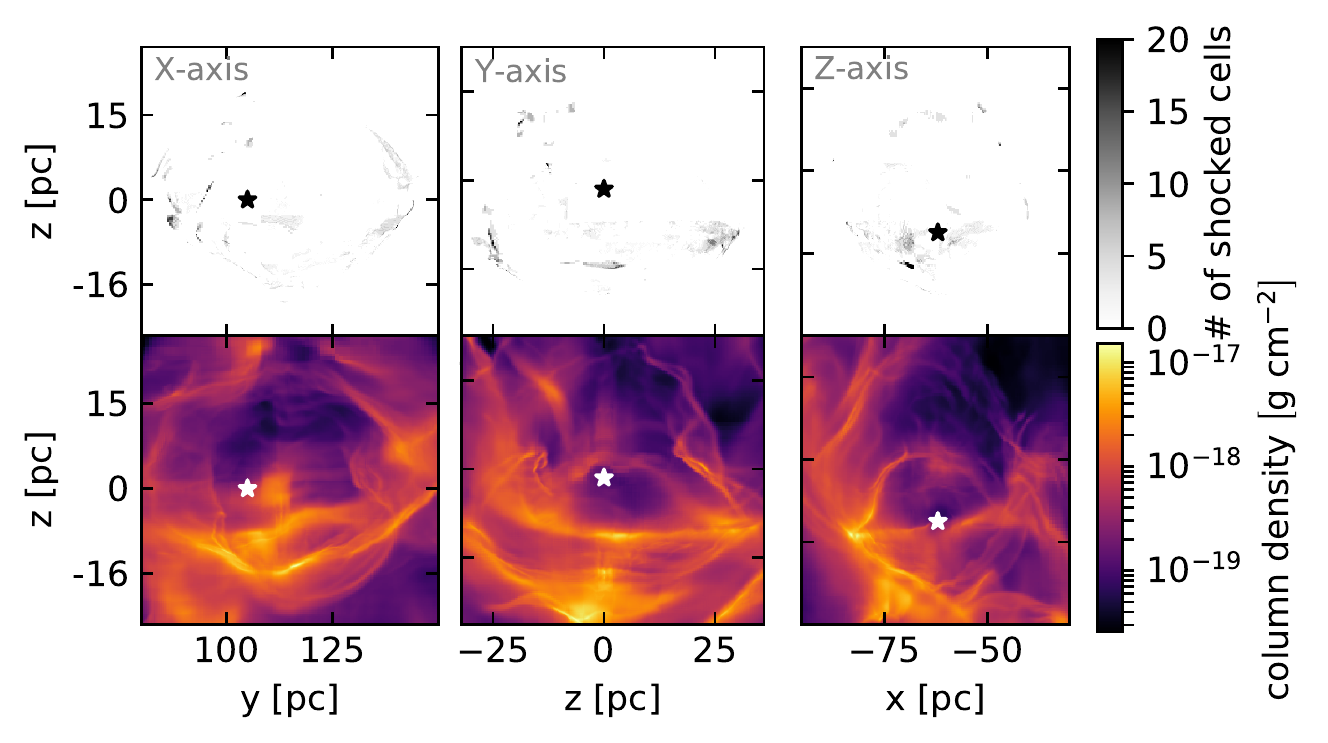}
    \caption{\textit{Top row}: projection of the shock cells along three the z, y and x for time 0.13 Myr. \textit{Bottom row}: column density, same parameters as for the top plot. The star symbol shows the position of the SN explosion.}
    \label{fig:shock_tau_ha}
\end{figure*}

\begin{figure*}
	\includegraphics[width=14cm]{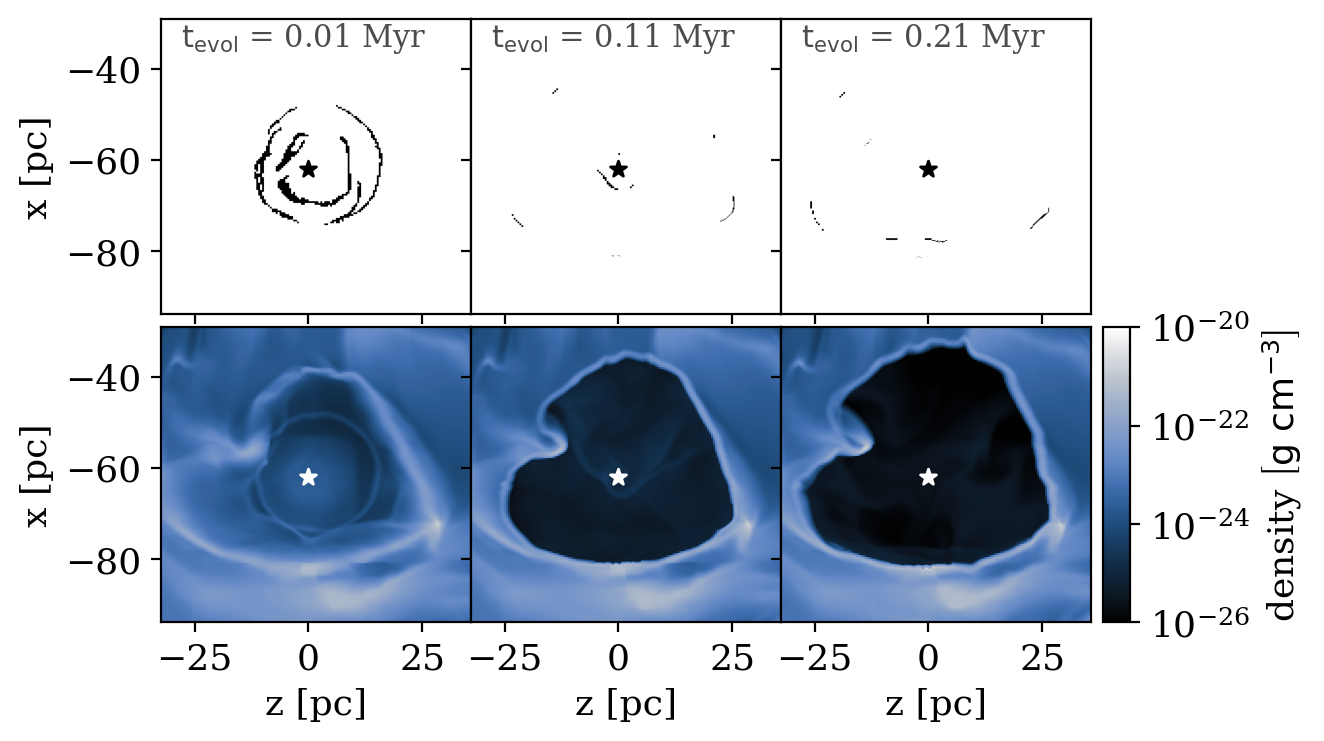}
    \caption{\textit{Top row}: time evolution from left to right of the position of shocked cells (forward and reverse shock) on the slice. \textit{Bottom row}: slice density.}
    \label{fig:shock_cells_dens}
\end{figure*}

This section is intended to show the position of the shock waves in the 2D and 3D cases. The first Fig.~\ref{fig:shock_tau_ha} shows the projections of the supernova bubble along three different axes for the cells where the shock was detected (top row) and the density projection (bottom row). Due to the increase in colour from white to black, you can see the overlap of some areas of the bubble (due to the asymmetry) and the strengthening of the shock waves. However, it is difficult to see this in the density projection Fig.~. The second Fig.~ \ref{fig:shock_cells_dens} shows the section of the supernova bubble in cells with detected shock waves and their time evolution (from left to right). Here you can see the initial shock wave from the centre and the reverse shock wave, which has a very complex shape due to the nonuniform density structure of the surrounding media.

\bsp	
\label{lastpage}
\end{document}